\documentclass[11pt]{article}
\usepackage{amsfonts}

\usepackage{amssymb}
\usepackage{amsmath}
\usepackage{epsfig}

\begin{document}
\newtheorem{theorem}{Theorem}
\newtheorem{corollary}{Corollary}
\newtheorem{conjecture}{Conjecture}
\newtheorem{definition}{Definition}
\newtheorem{lemma}{Lemma}

\newcommand{\define}{\stackrel{\triangle}{=}}

\pagestyle{empty}

\def\QED{\mbox{\rule[0pt]{1.5ex}{1.5ex}}}
\def\proof{\noindent{\it Proof: }}

\date{}

\title{Aiming Perfectly in the Dark - Blind Interference Alignment through Staggered Antenna Switching}
\author{\normalsize  Chenwei Wang, Tiangao Gou, Syed A. Jafar\\
      {\small \it E-mail~:~\{chenweiw,tgou,syed\}@uci.edu} \\
       }
\maketitle

\thispagestyle{empty}
\begin{abstract}
We propose a blind interference alignment scheme for the vector
broadcast channel where the transmitter is equipped with $M$
antennas and there are $K$ receivers, each equipped with a
reconfigurable antenna capable of switching among $M$ preset modes.
Without any knowledge of the channel coefficient values at the
transmitters and with only mild assumptions on the channel coherence
structure we show that $\frac{MK}{M+K-1}$ degrees of freedom are
achievable. The key to the blind interference alignment scheme is
the ability of the receivers to switch between reconfigurable
antenna modes to create short term channel fluctuation patterns that
are exploited by the transmitter. The achievable scheme does not
require cooperation between transmit antennas and is therefore
applicable to the $M\times K$ $X$ network as well. Only finite
symbol extensions are used, and no channel knowledge at the
receivers is required to null the interference.
\end{abstract}
\newpage

\section{Introduction}

Bandwidth is a precious resource for wireless networks. How to
share this limited resource among multiple users is the  primary
challenge. A new signal multiplexing approach, called
interference alignment, has shown recently that the bandwidth
available to each user can be \emph{greatly} improved
\cite{Cadambe_Jafar_int,Nazer_ergodic,
Jafar_Ergodic}. However, there remain very substantial hurdles in
translating existing interference alignment schemes to practice. The
most significant of these is the issue of channel knowledge. With
few exceptions, the interference alignment schemes proposed so far
require perfect, and often global, channel knowledge. In a network
of distributed nodes, with time-varying channel coefficient values,
this is very difficult if not altogether impossible. With this
challenge as our motivation, in this work we pursue the goal of
blind interference alignment, i.e., we seek to align interference
with no knowledge of channel conditions at either transmitters or
receivers.

Interference alignment predicates a design of signals so that they
cast overlapping shadows at the receivers where they constitute
interference while they remain distinguishable at the receivers
where they are desired \cite{Jafar_Shamai}. Finding ways to
simultaneously satisfy these seemingly conflicting requirements on
signal design has been the principal challenge in interference
alignment research. Existing works on interference alignment schemes
have looked for increasingly sophisticated mechanisms for aligning
interference, especially as the number of users increases. The most
powerful state-of-art interference alignment techniques draw upon
esoteric results from diophantine approximation theory, e.g., the
separability of lattices scaled by rationally independent scalars Ð
to fashion interference alignment in seemingly over-constrained
settings \cite{Etkin_Ordentlich, Real_IA, Gou_Jafar_Wang}.
Unfortunately, these powerful results also become increasingly
fragile in their absolute reliance on infinite precision channel
knowledge. The simplest setting where both the sophistication and
fragility of state-of-art interference alignment schemes is
simultaneously evident is the MISO broadcast channel (BC) with
multiple multicasts\cite{Gou_Jafar_Wang}, better known as the
compound MISO BC.

\begin{figure}[!h] \centering
\includegraphics[width=3.0in]{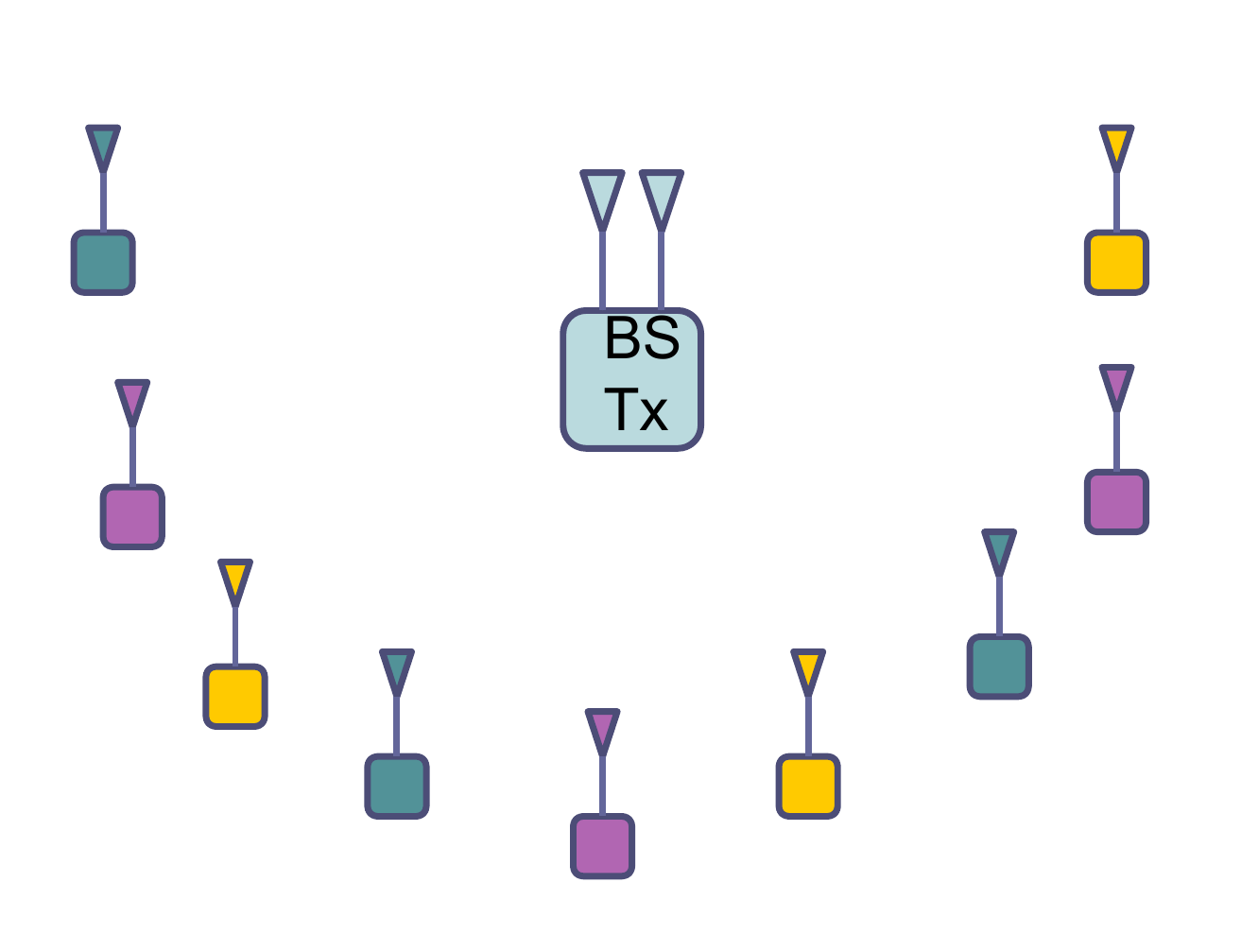}
\caption{Vector broadcast channel with $M=2$ antennas at the base
station and $K=3$ multicast groups (colors) of single-antenna
receivers, each group containing $J=3$ receivers}
\label{fig:misomulticast}
\end{figure}

A MISO broadcast channel with multiple multicasts, or the compound
MISO BC, refers to the setting where the base station transmitter is
equipped with $M$ transmit antennas and each receiver is equipped
with a single receive antenna. There are $K$ independent messages
intended for $K$ groups of receivers, each group consisting of  $J$
distinct receivers who want the same message. The channel states of
all users are globally and perfectly known to the transmitters and
recievers. This setting is interesting in its own right as a model
for tactical communications and also for commercial applications
where, say, mobile subscribers wish to watch one out of $K$
different live sports channels, so that each channel is being
watched by multiple users. From the perspective of channel
uncertainty, it also models the conventional vector BC where we have
only $K$ receivers, and each receiver's channel may be in one of $J$
different states. While the transmitter knows the $J$ possible
states, it does not know which one of these states is the actual
realization of the users' channel, and so it must encode the users'
data so that it can be reliably decoded in \emph{each} of the $J$
possible states of that user.

The study of the degrees of freedom for the compound MISO BC was
pioneered by Weingarten, Shamai and Kramer in
\cite{Weingarten_Shamai_Kramer}. In this work, it was shown that the
multiplexing gain in this setting cannot be more than
$\frac{MK}{M+K-1}$. It was also conjectured that as the number of
users $J$ in each group increases, the multiplexing gain will
collapse to unity. Recent work in \cite{Gou_Jafar_Wang,
Maddah_Ali} disproved this conjecture and showed that
interference alignment is still feasible in the compound vector
broadcast channel so that a total multiplexing gain of
$\frac{MK}{M+K-1}$ is achievable regardless of the number users $J$
in each group, provided $J$ is finite. The surprising result  is
based on an interference alignment scheme that exploits the
separation properties of lattices scaled by rationally independent
scalars. Rational independence of channel coefficients is a
meaningful concept only when they are known to infinite precision.
Thus,  the alignment scheme is both powerful in its ability to align
interference regardless of the number of users $J$ in each group,
and also fragile in its fundamental reliance upon the assumption
that the transmitter knows all the channel states with infinite
precision.


The compound MISO BC with increasing number of alignment constraints
(as $J$ increases) is analogous to a marksmans' challenge where one
must hit an increasing number of targets with each shot. While the
previous conjecture  in \cite{Weingarten_Shamai_Kramer} considered
this challenge to be impossible as the number of targets increases,
the new result of \cite{Gou_Jafar_Wang} shows that an extreme
marksman (the interference alignment scheme over rational
dimensions) is indeed capable of meeting the challenge. However, as
one might expect, it turns out that extreme marksmanship requires
perfect eyesight -- in this case, the ability to separate numbers
into rational dimensions with infinite precision.

Continuing the same line of thought, the goal of blind interference
alignment, much like aiming perfectly in the dark, may appear
impossible.  Yet, it has been shown in \cite{Jafar_SCBC} that in
certain scenarios blind interference alignment is not only possible,
but also it can be quite simple.  For instance, in the context of
the MISO BC with $M=2$ antennas at the base station and $K=2$ users,
even when the transmitter has no CSIT and the channel coefficients
may be drawn from a continuum, \cite{Jafar_SCBC} showed that if the
coherence blocks of the two users are suitably staggered then the
outer bound value of $\frac{4}{3}$ DoF  is achievable. Moreover, the
achievable scheme is a form of repetition coding over a
\emph{supersymbol} consisting of three channel uses.

The supersymbol structure used for blind interference alignment in
\cite{Jafar_SCBC} arises naturally in certain practical settings,
e.g., if user 1 has a larger coherence time than user 2, while user
2 has a larger coherence bandwidth than user 1. However, in general
, e.g., if the channel coherence does not display any special
properties, the problem of blind interference alignment remains
open. In this work we address this setting and consider a bolder
approach, summarized in the following question --\emph{can we
artificially manipulate the channel itself to create the
opportunities that facilitate blind interference alignment?}

\begin{figure}[!h]
\centering
\includegraphics[width=3.0in]{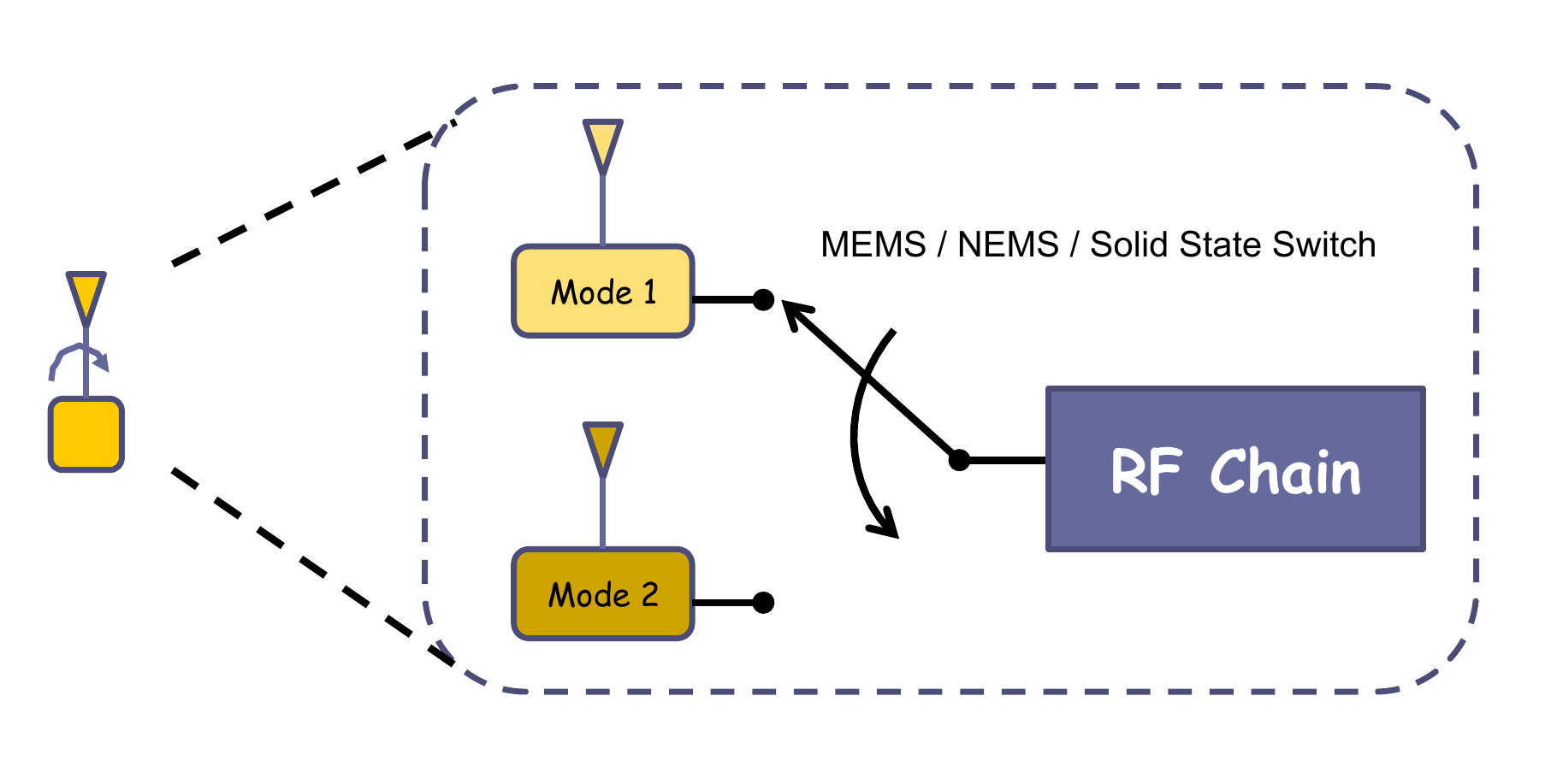}
\caption{Reconfigurable antenna - Conceptual depiction}
\label{fig:mra}
\end{figure}

The goal of manipulating the channel naturally leads us to
reconfigurable antennas. A reconfigurable antenna is an antenna that
can change its characteristics by dynamically changing its geometry.
Specifically, the current distribution over the volume of the
antenna is changed by switching on and off various geometrical
metallic segments (pixels, strips etc.) that constitute the
reconfigurable antenna. Each distinct geometrical configuration
corresponds to a different mode of operation. Various technologies,
such as microelectromechanical switches (MEMS),
nanoelectromechanical switches (NEMS) or solid state switches are
used to perform the switching operation and offer various tradeoffs
in terms of the switching speed, insertion loss, monolithic
integration capability, size and reliability\cite{Bedri,Mehta,Rebeiz}. A
reconfigurable antenna offers a choice to switch among several
pre-set modes. It is more flexible than a single conventional
antenna in its ability to switch its radiation pattern among a fixed number of
\emph{ preset modes}, and yet it is less flexible
than multiple conventional antennas which can be used together with
arbitrary beamforming weights to construct logical beams over a
continuum of possibilities. For many applications, a single
reconfigurable antenna is more desirable than multiple conventional
antennas because (1) it needs only one RF chain, and (2) the
integrated design of a reconfigurable antenna makes it smaller than
multiple conventional antennas. However, the popularity of
reconfigurable antennas has been limited by the assumption of fixed
\emph{preset} modes which do not allow continuous adaptability
needed for beamforming techniques such as zero forcing, that are
typically used to multiplex signals. As a consequence,
reconfigurable antennas have heretofore been explored for diversity
benefits, but not for multiplexing gain. The focus of research in
reconfigurable antenna fabrication has been to  create desired
radiation patterns. However, in this work we use antenna switching
not to direct the beam in a specific direction, but rather to
introduce channel fluctuations at pre-determined time instants. More
importantly, we will use \emph{blind} antenna switching, i.e., the
antenna switching will not be based on the channel state information
at the receivers (CSIR). This relaxed requirement could also
simplify the design of the reconfigurable antenna. For our purpose
in this work, we ignore the specific hardware aspects and focus
instead on the conceptual model of a reconfigurable antenna. At a
conceptual level, we model a reconfigurable antenna as capable of
switching among independent dumb (isotropic) modes, shown in Fig.
\ref{fig:mra}. Note that the conceptual model is identical to
antenna selection.

Consider again the $M=K=2$ multicast setting. Suppose all the
channels are statistically equivalent, follow the same temporal
correlation model, and there is no CSIT. This is a worst case
scenario and no multiplexing of signals is possible. The total DoF =
1 and time-division, i.e., serving only one group of users at a
time, is optimal. Now, let us bring reconfigurable antennas or
antenna switching into the picture, so each receiver can switch
between two antenna modes. The traditional approach for antenna
switching is for each receiver to choose the mode that allows the
greatest signal strength for that particular receiver\cite{Molisch_Win,Sanayei_Nostratinia}. We call this
``selfish" antenna switching. Since all receive antennas are
statistically equivalent and they follow the same selfish antenna
switching policy, and since the transmitter has no instantaneous
CSIT, it is easily seen that the users remain statistically
equivalent. In other words, the best the transmitter can do is to
transmit to one user at a time and use orthogonal time division
across users, so that the total DoF cannot be more than unity. The
conventional (selfish) antenna selection approach therefore does not
allow any DoF advantage.

\begin{figure}[!t]
\centering
\includegraphics[width=4.8in, trim=0 20 0 100]{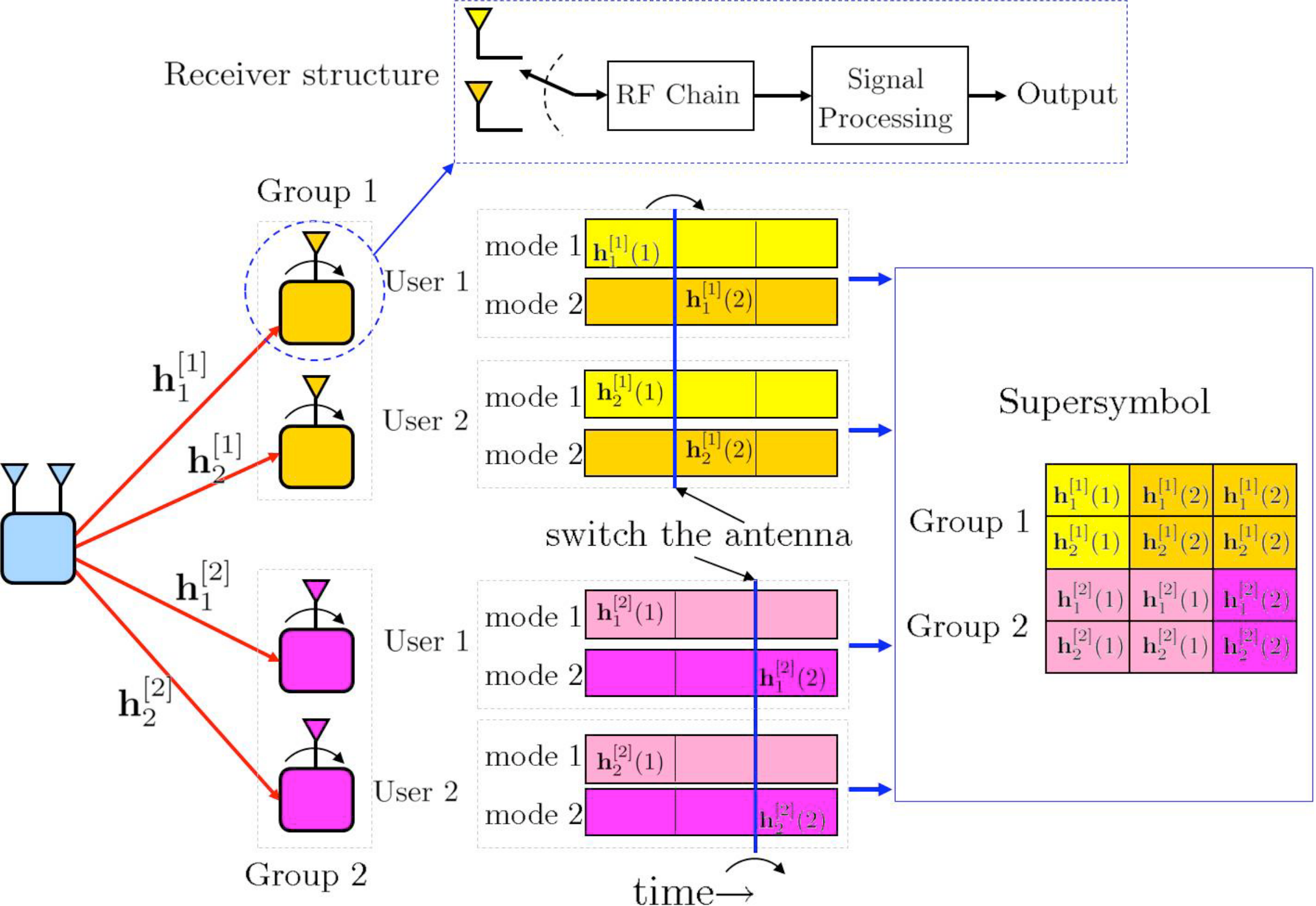}
\caption{MISO BC with Blind Staggered Antenna Switching and the Resulting Supersymbol Structure}
\label{fig:antennaselection}
\end{figure}

The following insight is distilled from \cite{Jafar_SCBC} and forms
the basis for this work - ``\emph{Instead of each receiver
(selfishly) selecting the best reconfigurable antenna mode, if the
receivers blindly switch antennas according to a pre-determined
signature pattern, then they can artificially create the staggered
coherence block structures needed for blind interference
alignment}".

Continuing with the $M=K=2$ example, suppose each receiver blindly
switches from one preset mode to another every two symbols. However
the switching instants are staggered. While all receivers in group 1
switch on odd time slots, the receivers in group 2 switch on even
time slots. This creates the supersymbol structure shown in Figure
\ref{fig:antennaselection} which is the key to the  blind
interference alignment scheme in \cite{Jafar_SCBC}. Each supersymbol
consists of 3 symbols. The channel state of every user in one group
(labeled as user 1 in the figure) changes after the first symbol and
remains fixed for the last 2 symbols, while the channel state of
every user in the other group (user 2 in the figure) is fixed for
the first 2 symbols and changes in the last symbol. Once this
supersymbol structure is achieved, the staggered coherence block
coding (SCBC) scheme proposed in \cite{Jafar_SCBC} is directly
applied to achieve $\frac{4}{3}$ DoF, regardless of the number of
receivers in each group. Further, note that this is accomplished
without any special assumptions on the channel coherence block
structures, except that the coherence time for each user is long
enough to span the coding block.

The $M=2, K=2$ MISO BC example suggests that staggered antenna
switching (SAS) with reconfigurable antennas can be used to
translate the staggered coherence block coding schemes of
\cite{Jafar_SCBC} into practice. However, there are important
distinctions in terms of what kinds of supersymbol structures can
result in the setting of \cite{Jafar_SCBC} versus our setting in
this paper. For example, consider the $K$ user interference channel
for which an SCBC scheme is proposed in \cite{Jafar_SCBC} that is
capable of achieving $K/2$ DoF. This SCBC scheme is contingent on a
very specific temporal correlation model -- the cross channels
follow the same temporal correlation structure which is different
from the direct channels. However, note that antenna switching at a
receiver cannot selectively change the channels for \emph{some}
transmitters while keeping the channels from \emph{other}
transmitters to that same receiver unchanged. When we switch a
receive antenna, \emph{all} the channel coefficients associated with
that receive antenna will change. Thus, some supersymbol structures
possible with the staggered coherent block model of
\cite{Jafar_SCBC} are not possible with staggered antenna switching
of reconfigurable antennas. On the other hand, the converse is also
true. Staggered antenna switching, because it can follow any complex
pattern, allows some supersymbol structures that cannot result
simply from staggered coherent blocks. Thus, the staggered antenna
switching (SAS) framework that we explore in this paper has a
distinct character from the staggered coherence block coding (SCBC)
framework considered in \cite{Jafar_SCBC} and in general neither can
be seen as a special case of the other.

We summarize the salient features of the blind interference alignment scheme proposed in this work as follows.

\begin{enumerate}
\item No CSIT is required to align interference. Unlike \cite{Weingarten_Shamai_Kramer, Gou_Jafar_Wang} the channel uncertainty may be spread over a continuum.
\item No cooperation among transmit antennas is needed. Thus the BC results extend to the X channel setting. This is similar to the result in \cite{Gou_Jafar_Wang, Jafar_SCBC}.
\item The multiplexing gain achieved with no CSIT for the MISO BC with multiple ($J$) multicasts is the same as the maximum possible multiplexing gain with full CSIT for large $J$.
\item The multiplexing gain achieved with no CSIT for the X channel, is the same as the maximum possible multiplexing gain with full CSIT. This is similar to the result in \cite{Gou_Jafar_Wang}. 
\item Alignment is achieved by coding over only a finite number of symbols. This is in contrast to \cite{Cadambe_Jafar_X} where infinite symbol extensions are needed to achieve the outer bound for X channel even with perfect CSIT.
\item No CSIR is needed for the antenna switching pattern used by each receiver.
\item No CSIR is required to null the interference (without losing the desired signal dimensions). While we do not consider non-coherent communication in this paper, the proposed scheme can be  directly extended to non-coherent settings with e.g. differential coding schemes.
\item Unlike \cite{Jafar_SCBC} no special assumptions are needed on channel coherence structure. In fact, except for the antenna switching patterns, the receivers may even be statistically equivalent, i.e., indistinguishable to the transmitter.
\end{enumerate}
\section{System model}\label{sec:systemmodel}
Consider, as before, the MISO BC with multiple multicasts, where the
transmitter has $M$ antennas while each receiver is equipped with
one reconfigurable antenna (and thus only one RF chain) that can
switch among $M$ preset modes.  Since there is no CSIT, the multiple
multicast setting is subsumed under the conventional MISO BC setting
where we have a unique receiver for each independent message. We
will focus on this latter scenario. The results can be extended
easily to the multiple multicast setting by having every receiver in
a particular group follow the same strategy as the unique receiver
that stands as proxy for this group in the classical MISO BC.

Let $K$ be the number of receivers. Further, let us denote the
$1\times M$ channel vector associated with the $m^{th}$ preset mode
of user $k$'s reconfigurable receive antenna as
$\mathbf{h}^{[k]}(m)\in\mathbb{C}^{1\times M}$ where
$k\in\mathcal{K}=\{1,2,\cdots,K\}$ and $m\in
\mathcal{M}=\{1,2,\cdots,M\}$. We assume that the channel vectors
are generic, by which we mean that they are drawn from a continuous
distribution (bounded away from zero and infinity to avoid
degenerate cases), so that any $M$ of them are linearly independent
almost surely.

We make no special assumption on the channel coherence block
structures, except that the coherence times are long enough so that
the channels stay constant across a supersymbol. The supersymbols
will be defined later for each $M,K$.

With the staggered antenna switching scheme, the receivers switch
between their antenna modes in a predetermined pattern, so that at
time $t$, the mode selected by receiver $k$ is $m^{[k]}(t)$, and the
resulting channel for user $k$ is represented as
${\mathbf{h}}^{[k]}(m^{[k]}(t))$. Thus, the received signal for the
$k^{th}$ user, at time $t$, is
\begin{eqnarray}
y^{[k]}(t)= {\mathbf{h}}^{[k]}(m^{[k]}(t)){\bf x}(t)+z^{[k]}(t)~~~~k\in\mathcal{K}, m^{[k]}(t)\in\mathcal{M}
\end{eqnarray}
where $\mathbf{x}(t)\in\mathbb{C}^{M\times 1}$ is an $M \times 1$
transmitted signal vector  and $z^{[k]}(t)\sim\mathcal{CN}(0,1)$ is
the additive white Gaussian noise (AWGN). The channel input is
subject to an average power constraint
$\mathbb{E}[\|\mathbf{x}\|^2]\leqslant P$. Unless explicitly stated
otherwise, we assume no channel state information at the transmitter
(CSIT), i.e., the channel coefficient values are not known to the
transmitters. While the blind interference alignment scheme does not
require channel state information at the receivers either, i.e., we
need no CSIR to either align interference or to null it out at the
receiver,  we will keep the assumption of perfect CSIR primarily in
order to be able to define the degrees of freedom (or multiplexing
gain) as the capacity pre-log, and thereby provide us a clean metric
for gauging the extent of interference alignment. We will point out
the no-CSIR requirement feature of our proposed scheme later on in
this paper. While the channel coefficient values are not known, we
assume that the switching pattern functions $m^{[k]}(t)$, because
they are pre-determined by design (like the codebooks), are known to
everyone.

The transmitter sends an independent message $W^{[k]}$ with rate
$R^{[k]}$ to receiver $k, \forall k\in\mathcal{K}$.  A rate tuple
${\bf R}=(R^{[1]},R^{[2]},\ldots, R^{[K]})$ is achievable if every
receiver  is able to decode its message with an error probability
that can be made arbitrarily small by coding over sufficient channel
uses. The closure of the set of all achievable rate tuples is the
capacity region $\mathcal{C}$. The degrees of freedom metric, $d$,
is defined as
\begin{eqnarray}
d=\lim_{P \rightarrow \infty}\max_{{\bf R}\in\mathcal{C}}\frac{R^{[1]}+\cdots+R^{[K]}}{\log P}
\end{eqnarray}

\section{Blind Interference Alignment for the $K$ User $M \times 1$ MISO BC}
We present the main result in the following theorem.

\begin{theorem}\label{theorem:kbcdof}
For the $K$ user $M\times 1$ MISO BC defined in Section
\ref{sec:systemmodel}, a total of $\frac{MK}{K+M-1}$ DoF are
achievable, almost surely.
\end{theorem}
As mentioned before, the achievable scheme relies on designing an
antenna switching pattern for each user and designing a beamforming
strategy based on the corresponding temporal correlation structure.
The goal is to achieve interference alignment, which refers to the
construction of signals in such a manner that they cast overlapping
shadows at the receivers where they constitute interference while
remain distinct where they are desired. In the following subsection,
we first highlight the key to the interference alignment schemes
used in this work.

\subsection{The Key to Blind Interference Alignment - The Alignment Block}
Consider the $K$ user $M \times 1$ MISO broadcast channel defined in
Section \ref{sec:systemmodel}. For simplicity, consider only the
transmission of the message for user 1. Suppose multiple symbols are
transmitted for user 1. These transmitted symbols for user 1 will
cause interference at all other users. Interference alignment means
that we would like to keep these symbols distinct at receiver 1, but
consolidate them into a smaller subspace at all other receivers.
Most importantly, we must accomplish this without any knowledge of
the channel coefficients.

\begin{figure}[!h]
\centering
\includegraphics[width=4.7in, clip=true, trim=0 170 0 100]{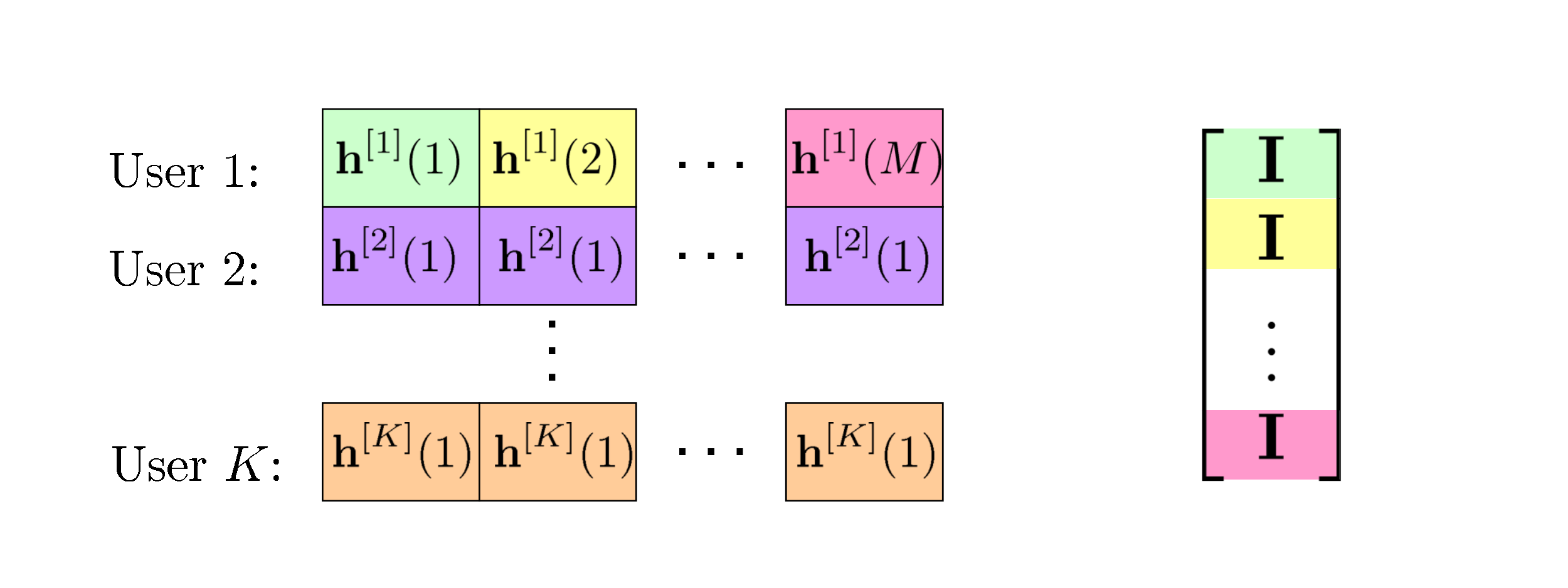}
\caption{Alignment block and the corresponding beamforming matrix
for $K$ user $M\times 1$ MISO BC} \label{fig:alignmentblock}
\end{figure}

{\it Staggered Antenna Switching Pattern:} Consider $M$ time slots.
During these $M$ time slots, receiver $1$ switches his
reconfigurable antenna mode each time to go through all $M$ modes.
All other receivers, $2,3\cdots, K$ do \emph{not} switch their
antenna modes, i.e., they listen to all $M$ transmissions through
the same channel. The resulting supersymbol structure is shown in
Figure \ref{fig:alignmentblock}.

{\it Beamforming:} During time slot $1$, suppose the transmitter
sends $M$ independent symbols for user $1$, one from each transmit
antenna. Now, let the transmitter repeat the same transmission over
$M$ time slots. In other words, transmit antenna $m$ repeats the
\emph{same} symbol $u^{[1]}_m$ a total of $M$ times, $\forall
m\in\mathcal{M}$. The transmitted vector can be represented as:

\begin{eqnarray}\label{eqn:beamformingmatrix}
\mathbf{X}=
\left[\begin{array}{c}
{\bf x}(1)\\
{\bf x}(2)\\
\vdots\\
{\bf x}(M)
\end{array}
\right]=
{\underbrace{\left[\begin{array}{c}
\mathbf{I}\\
\mathbf{I}\\
\vdots\\
\mathbf{I}
\end{array}\right]}_{\mbox{$M^2\times M$ beamforming matrix}}}
\left[
\begin{array}{c}
u^{[1]}_1\\
u^{[1]}_2\\
\vdots\\
u^{[1]}_M
\end{array}
\right]
\end{eqnarray}
where $\mathbf{I}$ is the $M \times M$ identity matrix. Note that the beamforming vectors do not depend on the channel values.


With this scheme (ignoring AWGN), the signal received at user 1 is
\begin{eqnarray*}\label{eqn:desiredsignal}
{\bf Y}=\left[\begin{array}{c}
y^{[1]}(1)\\
y^{[1]}(2)\\
\vdots\\
y^{[1]}(M)\end{array}\right]
&=&
\left[\begin{array}{cccc}
\mathbf{h}^{[1]}(1)&\mathbf{0}&\mathbf{0}&\mathbf{0}\\
\mathbf{0}&\mathbf{h}^{[1]}(2)&\mathbf{0}&\mathbf{0}\\
\mathbf{0}&\mathbf{0}&\ddots&\mathbf{0}\\
\mathbf{0}&\mathbf{0}&\mathbf{0}&\mathbf{h}^{[1]}(M)
\end{array}\right]_{M \times M^2}
\left[\begin{array}{c}
\mathbf{I}\\
\mathbf{I}\\
\vdots\\
\mathbf{I}
\end{array}\right]_{M^2\times M}
\left[
\begin{array}{c}
u^{[1]}_1\\
u^{[1]}_2\\
\vdots\\
u^{[1]}_M
\end{array}
\right]\\
&=&
{\underbrace{\left[\begin{array}{c}
\mathbf{h}^{[1]}(1)\\
\mathbf{h}^{[1]}(2)\\
\vdots\\
\mathbf{h}^{[1]}(M)
\end{array}\right]}_{M\times M \textrm{ channel matrix with rank}=M}}
\left[
\begin{array}{c}
u^{[1]}_1\\
u^{[1]}_2\\
\vdots\\
u^{[1]}_M
\end{array}
\right]
\end{eqnarray*}
where $\mathbf{h}^{[1]}(m)$, $m=1,\cdots, M$, is the $1\times M$
channel vector associated with the $m^{th}$ preset mode of user 1
and $\mathbf{0}$ is a $1\times M$  zero vector.  Since channels are
generic, user 1 accesses a full rank $M \times M$ MIMO channel
almost surely. Thus, the $M$ data streams cast an $M$-dimensional
shadow at receiver $1$, to achieve $M$ DoF.

Now consider the interference at the undesired users $k=2,\ldots,K$.
\begin{eqnarray}\label{eqn:interference}
\left[\begin{array}{cccc}
\mathbf{h}^{[k]}(1)&\mathbf{0}&\mathbf{0}&\mathbf{0}\\
\mathbf{0}&\mathbf{h}^{[k]}(1)&\mathbf{0}&\mathbf{0}\\
\mathbf{0}&\mathbf{0}&\ddots&\mathbf{0}\\
\mathbf{0}&\mathbf{0}&\mathbf{0}&\mathbf{h}^{[k]}(1)
\end{array}\right]_{M \times M^2}
\left[\begin{array}{c}
\mathbf{I}\\
\mathbf{I}\\
\vdots\\
\mathbf{I}
\end{array}\right]_{M^2\times M}
\left[
\begin{array}{c}
u^{[1]}_1\\
u^{[1]}_2\\
\vdots\\
u^{[1]}_M
\end{array}
\right]
=
\underbrace{\left[\begin{array}{c}
\mathbf{h}^{[k]}(1)\\
\mathbf{h}^{[k]}(1)\\
\vdots\\
\mathbf{h}^{[k]}(1)
\end{array}\right]}_{\textrm{rank}=1}
\left[
\begin{array}{c}
u^{[1]}_1\\
u^{[1]}_2\\
\vdots\\
u^{[1]}_M
\end{array}
\right]
\end{eqnarray}
Since all rows of the effective channel matrix in
\eqref{eqn:interference} are the same, its rank is equal to 1. Thus,
the $M$ symbols intended for user 1 cast only a $1$-dimensional
shadow on all undesired receivers, i.e., they align into one
dimension. Moreover, notice that they align along the $M\times 1$
vector $[1 ~1\cdots 1]^T$, regardless of the channel values.
Therefore, even if the receiver does not know the channel
coefficients, it can project the received signal into the null space
of the all-ones vector to zero force the interference. This is the
key to the blind interference cancellation at the receiver mentioned
earlier.

We summarize the key to blind interference alignment  as follows.
\emph{If the channel of the desired user changes while that of all
undesired users remains fixed over $M$ symbols, then the transmitter
can send $M$ data streams for the desired receiver without knowing
all channel values such that these beams remain distinguishable at
the desired user while they align into one dimension at all other
users}.

Note that with such an alignment scheme, $M$ interference vectors
can be aligned into one dimension. Thus, intuitively the achievable
$\frac{MK}{M+K-1}$ DoF for $K$ user $M \times 1$ MISO BC can be interpreted
as follows -- Every receiver demands $M$ DoF for a total of $MK$
DoF; at each receiver, the desired signals occupy $M$ dimensions
while $M(K-1)$ interference streams are aligned into $K-1$
dimensions, for a total of $M+K-1$ dimensions.

The supersymbol structure shown in Figure \ref{fig:alignmentblock}
will serve as the building block for designing the supersymbol,
i.e., antenna switching patterns,  for the general $K$ user $M\times 1$ MISO
BC. Thus, we refer to it as the \emph{alignment block} in the
following part of this paper. Our goal is to construct the alignment
block for each user. Furthermore, with the alignment block, the
design of beamforming vectors becomes straightforward. As shown in
Figure \ref{fig:alignmentblock}, over the $M$ symbols of the
alignment block, the beamforming matrix is obtained by stacking the
$M \times M$ identity matrix $M$ times. Finally, the $M$ symbols
constituting the alignment block may not be necessarily consecutive.
In this case, it can be obtained through an interleaving of symbols.

With this insight, we begin with the $K$ user $2 \times 1$ MISO BC, go on to
$K$ user $3 \times 1$ case , and at last solve the general $K$ user $M \times 1$ case.

\subsection{$K$ User $2\times 1$ MISO Broadcast Channel}
In this section, we consider the $K$ user $2 \times 1$ MISO BC where
the transmitter is equipped with two antennas and each of $K$
receivers has one reconfigurable antenna capable of switching
between 2 preset modes. We will show a total of $\frac{2K}{K+1}$ DoF
can be achieved. We begin with the 2 user case. Although the
solution for this case, as mentioned in the introduction, follows
directly from \cite{Jafar_SCBC}, here we use this simplest setting
as an example to illustrate how to use the alignment  block
mentioned in the last section to construct the supersymbol
structure. Note that the alignment block for $K$ user $2 \times 1$ case
consists of two symbols, over which the channel state changes at the
desired user while remains fixed at all undesired users.

\begin{figure}[!t]
\centering
\includegraphics[width=4.7in, trim=0 150 0 100]{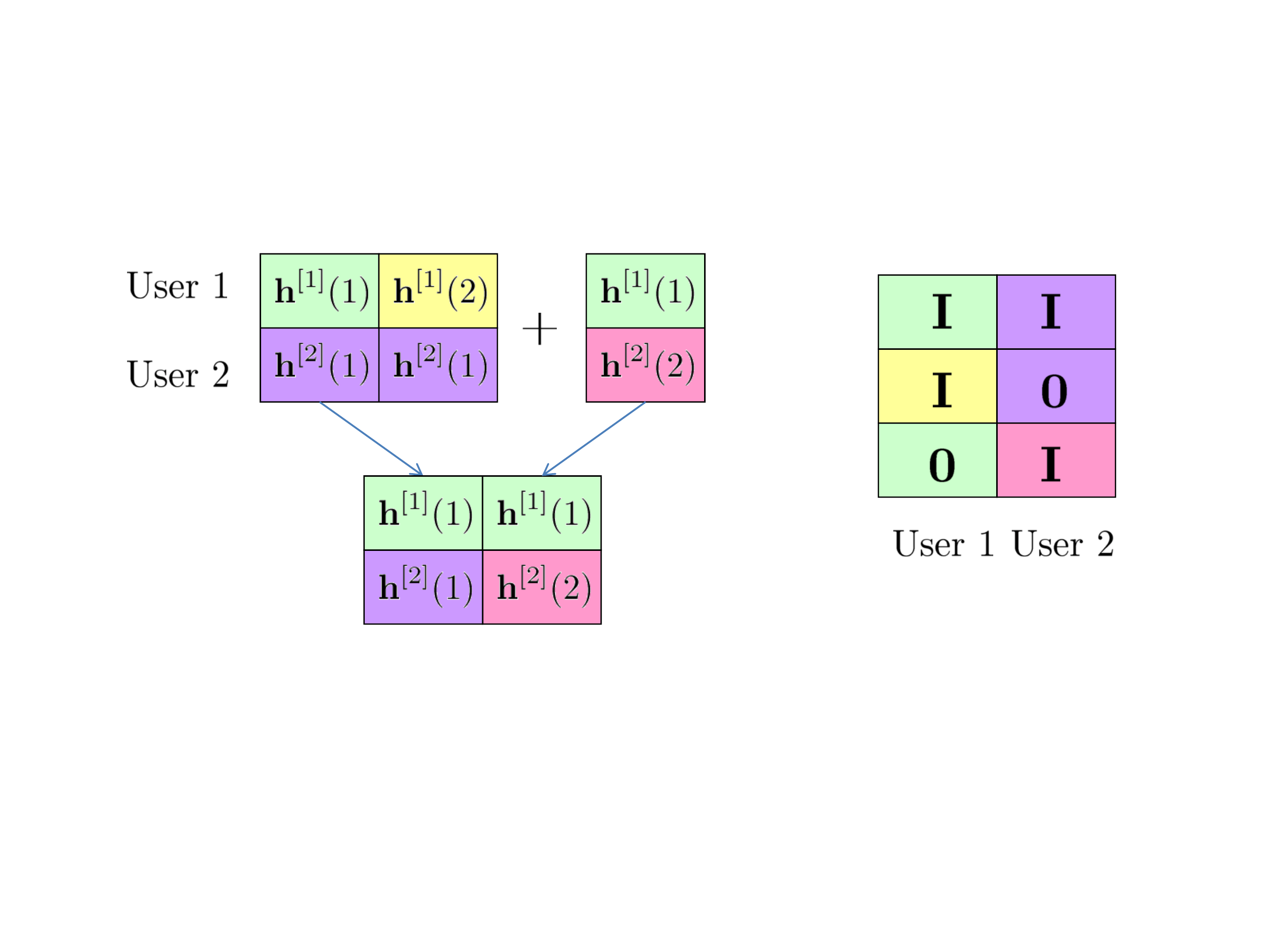}
\caption{The supersymbol structure for $2$ user $2\times 1$ MISO BC}
\label{fig:2by2supersymbol}
\end{figure}

For the 2 user $2\times 1$ MISO BC, our goal is to achieve
$\frac{4}{3}$ DoF. This can be done by sending two data streams,
each carrying one DoF, to each user over three symbol extensions. As
a result, in the three dimensional signal space of each user, the
desired signals occupy two dimensions, leaving one dimension for the
interference. Recall that over one alignment block consisting of two
symbols, two data streams can be sent to the desired user while
aligned into one dimension at the other user. Thus, if we can
construct two alignment blocks within three symbols, one for each
user, then interference can be aligned at both users. Moreover, if
at each user, the two dimensional space occupied by the desired
signals does not overlap with the space occupied by the
interference, then our objective is accomplished. With this
intuitive understanding, we design the supersymbol as shown in
Figure \ref{fig:2by2supersymbol}.
The first two symbols constitute
an alignment block for user 1. For user 2, we design the third
symbol such that through an interleaving of the first and third
symbols, they are converted into an alignment block.

Based on the supersymbol, we can design the $6 \times 2 $
beamforming matrix for each user as follows. The beamforming matrix
is constructed by stacking three $2 \times 2$ matrices, which is
equal to the number of symbols in the supersymbol. The $j^{th}$,
$j=1,2,3$, block in the beamforming matrix for each user corresponds
to the $j^{th}$ symbol in the supersymbol. For each user, if the
symbol belongs to his alignment block, then the corresponding block
in his beamforming matrix is a $2 \times 2$ identity matrix.
Otherwise, it is a $2\times2$ zero matrix. This is illustrated in
Figure \ref{fig:2by2supersymbol}. Since the first two symbols
constitute the alignment block for user 1, we place the identity
matrix at the first and second blocks but a zero matrix at the third
block. For user 2, the first and third symbols constitute an
alignment block; thus, we place the identity matrix at the first and
third blocks while a zero matrix at the second block.

Such a signaling scheme already guarantees the linear independence
of desired signals and the interference at receivers. Notice that at
the second time slot, the transmitter only transmits  to user 1 and
at the third time slot only to user 2. At user 1, the two desired
signal vectors do not occupy the third dimension (time slot) in the
three dimensional signal space while the interference vector
occupies it. Thus, the interference is linearly independent of the
desired signals. Similar argument can be applied to user 2. Next, we
formalize these ideas through mathematics.

The transmitted signal is
\begin{eqnarray}
\mathbf{X}=\left[\begin{array}{c}
\mathbf{I}\\
\mathbf{I}\\
\mathbf{0}
\end{array}\right]\left[\begin{array}{c}u^{[1]}_1\\ u^{[1]}_2\end{array}\right]+
\left[\begin{array}{c}
\mathbf{I}\\
\mathbf{0}\\
\mathbf{I}
\end{array}\right]\left[\begin{array}{c}u^{[2]}_1\\ u^{[2]}_2\end{array}\right]
\end{eqnarray}
where $\mathbf{I}$ is a $2\times 2$ identity matrix. $u^{[i]}_1,u^{[i]}_2, i=1,2$ are two independently encoded data streams intended to user $i$, each carrying one DoF.  With this scheme, the received signal at user 1 is
\begin{eqnarray}\label{eqn:rxsingalatuser1}
\underbrace{\left[\begin{array}{c}y^{[1]}(1)\\y^{[1]}(2)\\y^{[1]}(3)\end{array}\right]}_{\mathbf{y}^{[1]}}=\underbrace{\left[\begin{array}{c}
\mathbf{h}^{[1]}(1)\\
\mathbf{h}^{[1]}(2)\\
\mathbf{0}
\end{array}\right]}_{\textrm{rank}=2}\left[\begin{array}{c}u^{[1]}_1\\ u^{[1]}_2\end{array}\right]
+\underbrace{\left[\begin{array}{c}
\mathbf{h}^{[2]}(1)\\
\mathbf{0}\\
\mathbf{h}^{[2]}(1)
\end{array}\right]}_{\textrm{rank}=1}\left[\begin{array}{c}u^{[2]}_1\\ u^{[2]}_2\end{array}\right]+\underbrace{\left[\begin{array}{c}z^{[1]}(1)\\z^{[1]}(2)\\z^{[1]}(3)\end{array}\right]}_{\mathbf{z}^{[1]}}
\end{eqnarray}
where $\mathbf{0}$ is a $1 \times 2$ zero vector. From \eqref{eqn:rxsingalatuser1}, it can be easily seen that interference is aligned into one dimension along vector $[1~ 0~ 1]^T$ while the desired signals appear through a full rank matrix, and are therefore resolvable. In addition, notice that the third row of the desired signals is zero, while that of the interference vector, $[1~ 0~ 1]^T$, is 1. Therefore, any linear combination of the desired signals leads to zero in the third row, ensuring linear independence among them. Similar argument can be applied to user 2, so that he is able to achieve 2 DoF as well. Thus, 4 DoF can be achieved over 3 symbol extensions, so that $\frac{4}{3}$ (normalized) DoF are achieved. 

{\it Remark:} From equation (\ref{eqn:rxsingalatuser1}) note that user 1 can cancel the interference due to user 2 by simply subtracting the third received symbol from the first. This operation does not require any knowledge of channel coefficient values at the receiver and produces an interference-free signal while leaving the desired signal unaffected, although the noise is doubled over the first symbol. The blind interference cancellation property is common to all the blind interference alignment schemes proposed in this paper.

\begin{figure}[!t]
\centering
\includegraphics[width=5.0in, trim=0 110 0 100]{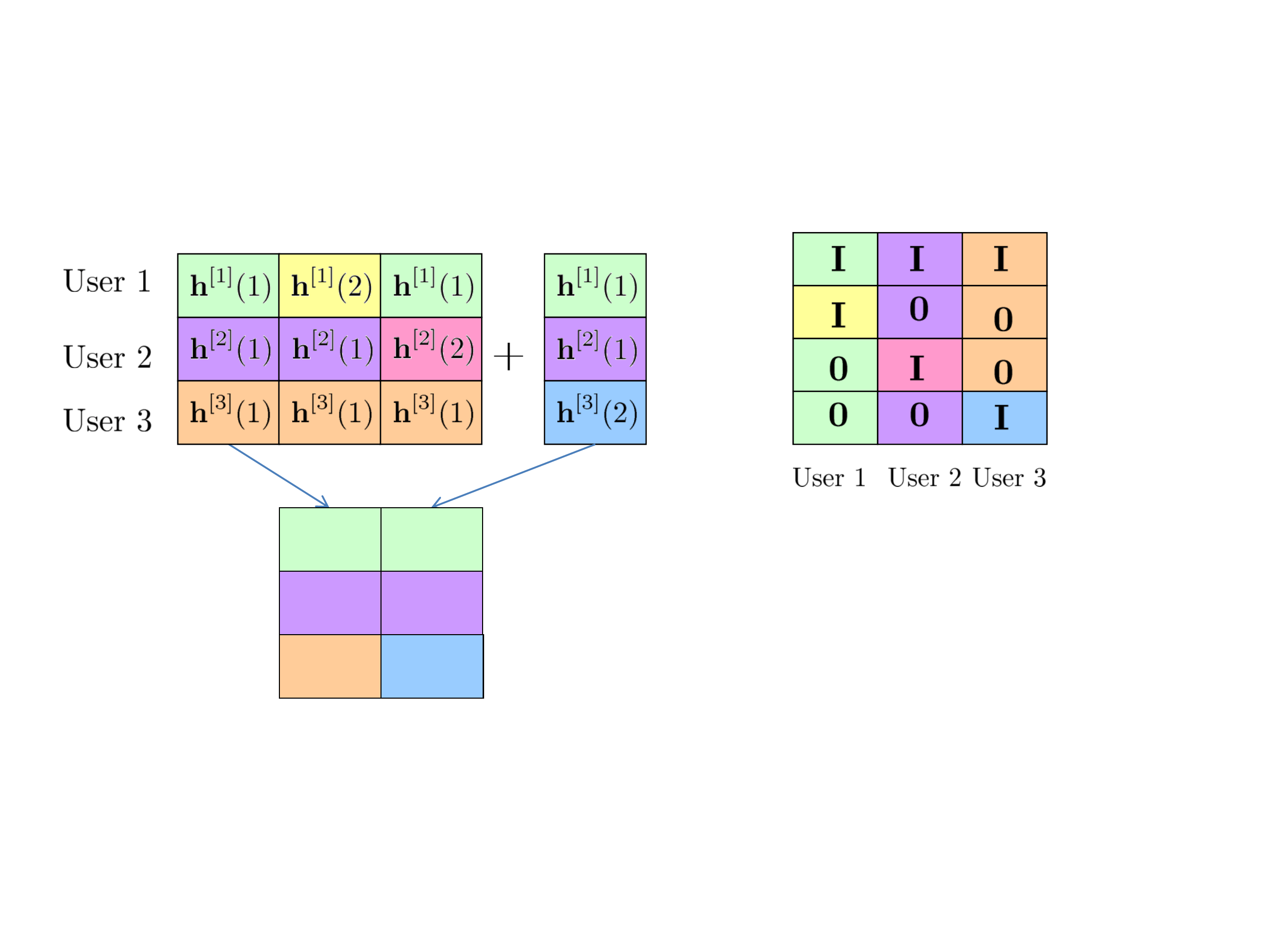}
\caption{The supersymbol structure for 3 user $2\times 1$ MISO BC}
\label{fig:2by3supersymbol}
\end{figure}

Now let us consider a 3 user $2 \times 1$ MISO BC.  For this case,
we need to show $\frac{3\times 2}{3+2-1}$ DoF can be achieved. This
can be done by achieving 2 DoF for each user over 4 symbol
extensions. Following similar analysis as two user case, we can
design the supersymbol consisting of 4 symbols such that three
alignment blocks, one for each user, are created. Then each user
transmits two data streams over its alignment block so that they
occupy two dimensions at the desired receiver's signal space while
aligning into one dimension at all other receivers. Thus, within
each user's four dimensional signal space, the desired signals
occupy two dimensions while the interference from each interferer
occupies one dimension for a total of two dimensions. The
supersymbol structure is shown in Figure \ref{fig:2by3supersymbol}.
Note that the first 3 symbols as the same as for the 2 user case.
The fourth symbol is designed such that when combined with the first
symbol, an alignment block is created for user 3. With the same
mapping from the supersymbol to the beamforming vectors as the two
user case,  the beamforming matrix is
\begin{eqnarray}
\left[
\begin{array}{ccc}
\mathbf{I}&\mathbf{I}&\mathbf{I}\\
\mathbf{I}&\mathbf{0}&\mathbf{0}\\
\mathbf{0}&\mathbf{I}&\mathbf{0}\\
\mathbf{0}&\mathbf{0}&\mathbf{I}\\
\end{array}
\right]
\end{eqnarray}
where $\mathbf{I}$ is the $2 \times 2$ identity matrix and
$\mathbf{0}$ is the $2 \times 2$ zero matrix. The $k^{th}$,
$k=1,2,3$, block column is for user $k$. With interference aligned,
it remains to check that desired signals do not overlap with the
interference at each receiver. Again, this is guaranteed by
orthogonality among signals over last three time slots, leading to
linear independence among them in the four dimensional signal space
at each receiver. To see this, let us consider user 1. The
interference from user 2 is aligned along vector $[1~0~1~0]^T$ and
the interference from user 3 is  aligned along the vector
$[1~0~0~1]^T$. Thus, we need to show that the following matrix is
full rank.
\begin{eqnarray}
\left[
\begin{array}{ccc}
\mathbf{h}^{[1]}(1)&1&1\\
\mathbf{h}^{[1]}(2)&0&0\\
\mathbf{0}&1&0\\
\mathbf{0}&0&1
\end{array}
\right]
\end{eqnarray}
This can be easily verified from the third and fourth rows. By
similar arguments, it can be shown that desired signals are linearly
independent with the interference at user 2 and 3.
\begin{figure}[!t]
\centering
\includegraphics[width=5.0in, trim=0 190 0 100]{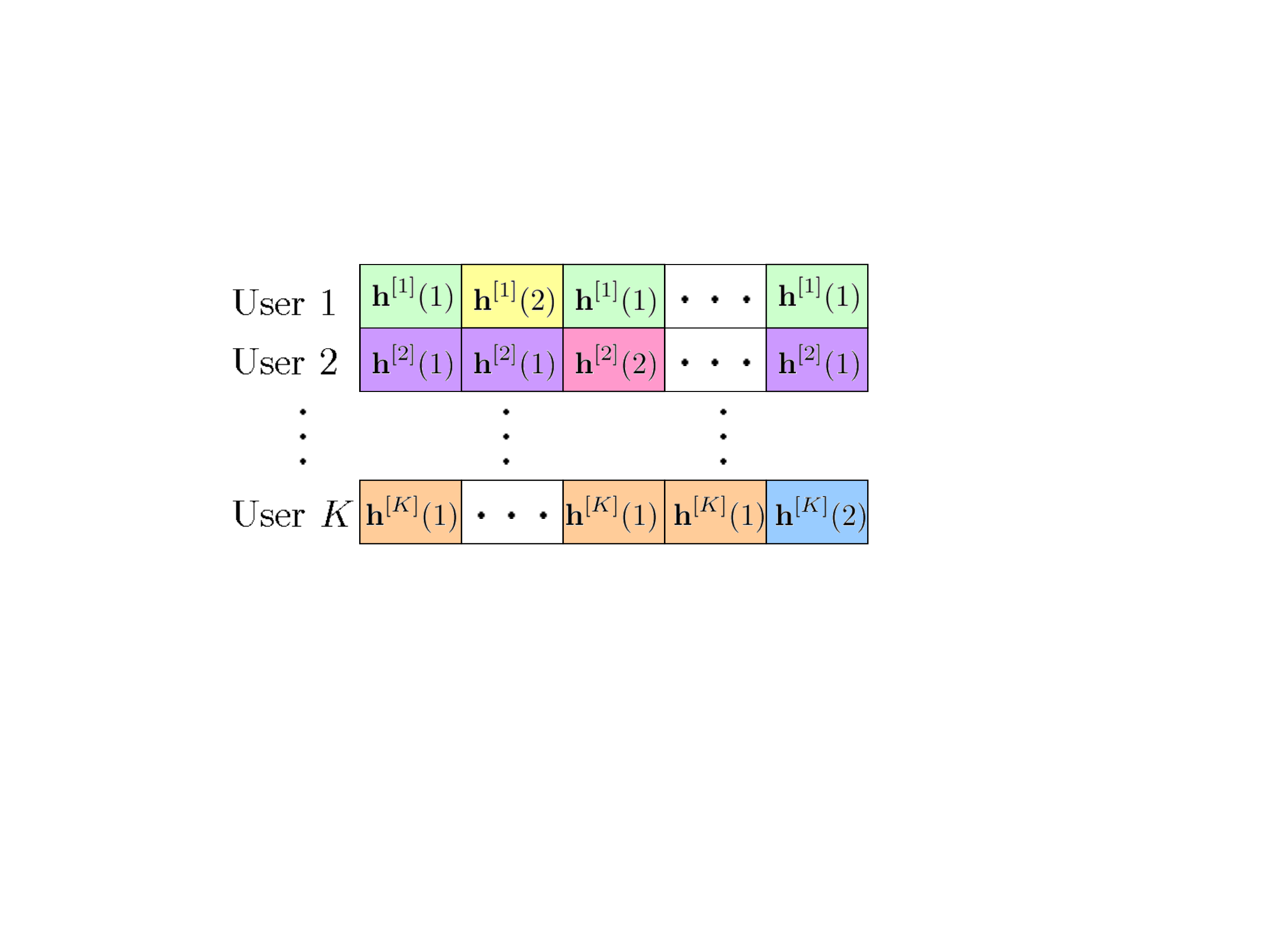}
\caption{The supersymbol structure for $K$ user $2\times 1$ MISO BC}
\label{fig:2byKsupersymbol}
\end{figure}

{\it Remark:} Once again, it is easily verified that each user can cancel all interference without the need for channel coefficient knowledge at the receivers. This is done, e.g., at user 1, by subtracting the third symbol from the first (to eliminate interference from user 2) and by subtracting the fourth symbol from the first (to eliminate interference from user 3), thus leaving only an interference-free desired signal  over symbols 1 and 2, albeit with the cost that the noise is now three times stronger for symbol 1. The property of blind interference cancellation at the receivers makes the blind interference alignment schemes in this work ideal for non-coherent communication, i.e., with no CSIR,  through differential coding schemes which only require coherence over relatively small intervals.

With the understanding of 2 user and 3 user cases, we are ready to
consider the general $K$ user $2 \times 1$ MISO BC. In general, the
supersymbol structure is shown in Figure \ref{fig:2byKsupersymbol}.
The supersymbol consists of $K+1$ symbols. Each user's channel state
only comes from one of two channel values corresponding to channels
associated with two reconfigurable modes of each receiver's
reconfigurable antenna. For user $k$, $k\in\mathcal{K}$, the channel
state over the first $k$ symbols maintains the first value, it
changes to the other value at the $(k+1)^{th}$ symbol and changes
back to the first value at the $(k+2)^{th}$ symbol and remains fixed
until the end of the supersymbol. With this structure, user $k$ can
obtain its alignment block through an interleaving of symbols at the
first time slot and the $(k+1)^{th}$ time slot. Over each alignment
block, two data streams, each carrying one DoF, are transmitted to
its corresponding user while remaining aligned into one dimension at
all other users. According to the supersymbol structure, the
beamforming matrix can be obtained as follows
\begin{eqnarray}\label{eqn:2byK}
\left[
\begin{array}{cccc}
\mathbf{I}&\mathbf{I}&\cdots&\mathbf{I}\\
\mathbf{I}&\mathbf{0}&\cdots&\mathbf{0}\\
\mathbf{0}&\mathbf{I}&\cdots&\mathbf{0}\\
\vdots&\vdots&\ddots&\vdots\\
\mathbf{0}&\mathbf{0}&\cdots&\mathbf{I}
\end{array}
\right]_{2(K+1) \times 2K}
\end{eqnarray}
The $i^{th}$ block column carries two DoF for user $i$. With this
scheme, at each user, the desired signals occupy 2 dimensions in the
$K+1$ dimensional signal space while all interference occupies $K-1$
dimensions, one from each interferer. It remains to show that at
each user, the desired signals and interference are linearly
independent. This is guaranteed by the orthogonality among signals
over the last $K$ symbols. At $(i+1)^{th}$, $i=1,\cdots, K$ time
slot, the transmitter only transmits signals to user $i$. Recall the
interference from user $i$ is aligned into one dimension at all
other users. Therefore, only this interference vector occupies the
$(i+1)^{th}$ dimension in the $K+1$ dimensional signal space,
leading to its linear independence with desired signals and all
other interference. Thus, at each user, all interference is linearly
independent with the desired signals.  As a result, each user can
achieve 2 DoF, for a total of $2K$ DoF, over $K+1$ symbol
extensions. Thus, $\frac{2K}{K+1}$ DoF are achieved.

\subsection{$K$ User $3 \times 1$ MISO BC}
When the number of antennas at the transmitter increases to 3, the
alignment problem is more challenging. However, the key idea, as
before, is to construct alignment blocks for each user. Note that
the alignment block for any user in the  $3 \times 1$ MISO BC is
comprised of three symbols, over which the that user's channel
changes while the channel of all other users remains fixed. Thus,
over the alignment block, three data streams can be separated at the
desired user while remaining aligned into one dimension at all other
users. We begin with the 2 user $3\times 1$ case.

\subsubsection{2 User $3 \times 1$ MISO BC}

\begin{figure}[!t]
\centering
\includegraphics[width=5.1in, trim=0 170 0 100]{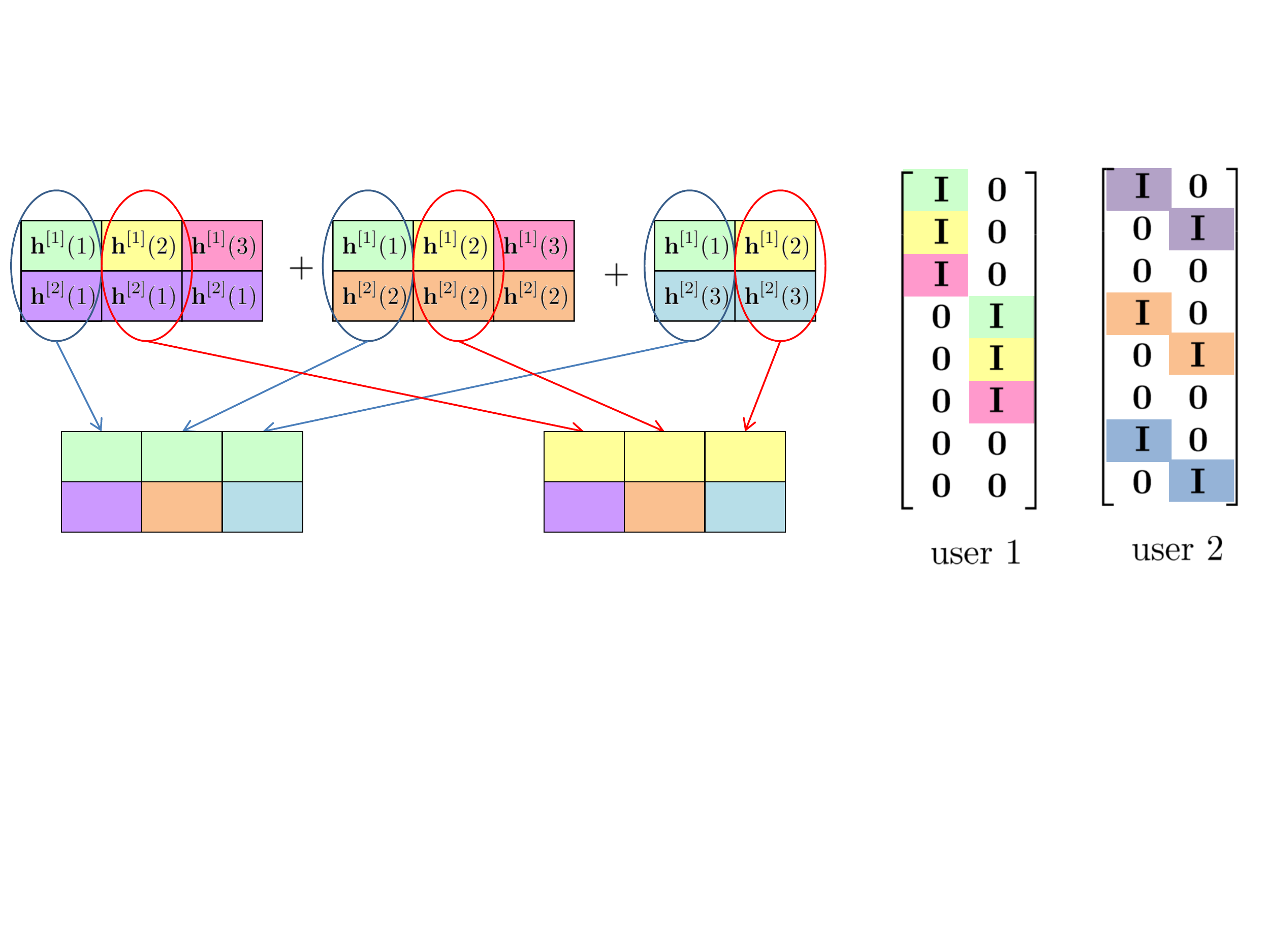}
\caption{The  supersymbol structure for 2 user $3\times 1$ MISO BC}
\label{fig:3by2supersymbol}
\end{figure}

We need to show $\frac{2\times3}{2+3-1}$ DoF are achievable. This
can be achieved with 8 symbol extensions, over which each user
achieves 6 DoF, for a total of 12 DoF. Each user's 6 beams can be
sent over two alignment blocks, 3 for each alignment block, so that
they can be aligned into two dimensions at the other receiver,
leaving the remaining 6 dimensions for the desired signals. The
supersymbol consisting of 8 symbols is shown in Figure
\ref{fig:3by2supersymbol}. As we can see, the first 6 symbols
constitute two alignment blocks for user 1. On the other hand, two
alignment blocks for user 2 can be obtained by adding 2 additional
symbols and through an interleaving of symbols as shown in Figure
\ref{fig:3by2supersymbol}. It is important to note that for each
user, two alignment blocks do \emph{not} overlap with each other.
Thus, signals sent over two alignment blocks are \emph{orthogonal}
in time. As a result, 3 beams sent over one alignment block are
orthogonal to the other 3 beams sent over the other alignment block,
ensuring 6 beams can be separated at the desired user.

With the supersymbol structure, we can design the beamforming
matrices. Similar to the 2 user case, the beamforming matrix for
each user can be obtained according to the alignment block.
Different from the 2 user case, where each user has only one
alignment block, here each user has two alignment blocks. In this
case, each user's beamforming matrix consists of two block columns,
each corresponding to one alignment block. Each block column can be
designed following the same mapping used in the two user case as
illustrated in Figure \ref{fig:3by2supersymbol}. With these
beamforming vectors, the transmitted signal is
\begin{eqnarray*}
\mathbf{X}=\underbrace{\left[\begin{array}{cc}
\mathbf{I}&\mathbf{0}\\
\mathbf{I}&\mathbf{0}\\
\mathbf{I}&\mathbf{0}\\
\mathbf{0}&\mathbf{I} \\
\mathbf{0}&\mathbf{I}\\
\mathbf{0}&\mathbf{I} \\
\mathbf{0}&\mathbf{0}\\
\mathbf{0}&\mathbf{0}
\end{array}\right]}_{\textrm{User 1}}
\left[\begin{array}{c}
u^{[1]}_1\\
u^{[1]}_2\\
\vdots\\
u^{[1]}_6
\end{array}\right]
+
\underbrace{\left[\begin{array}{cccc}
\mathbf{I}&\mathbf{0}\\
\mathbf{0}&\mathbf{I}\\
\mathbf{0}&\mathbf{0}\\
\mathbf{I}&\mathbf{0} \\
\mathbf{0}&\mathbf{I}\\
\mathbf{0}&\mathbf{0}\\
\mathbf{I}&\mathbf{0}\\
\mathbf{0}&\mathbf{I}
\end{array}\right]}_{\textrm{User 2}}
\left[\begin{array}{c}
u^{[2]}_1\\
u^{[2]}_2\\
\vdots\\
u^{[2]}_6
\end{array}\right]
\end{eqnarray*}
where $\mathbf{I}$ is the $3\times 3$ identity matrix and
$\mathbf{0}$ is a $3 \times 3$ zero matrix. $u^{[k]}_j$, $k=1,2$,
$j=1,\cdots, 6$, is the $j^{th}$ independently encoded data stream,
carrying one DoF, for user $k$. The channel matrix is a block
diagonal matrix with the $i^{th}$ diagonal block corresponding to
the $i^{th}$ symbol in the supersymbol. Thus, the channel matrix of
user 1 is
\begin{small}\begin{eqnarray*}
\mathbf{H}^{[1]}=\left[
\begin{array}{cccccccc}
\mathbf{h}^{[1]}(1)&\mathbf{0}&\mathbf{0}&\mathbf{0}&\mathbf{0}&\mathbf{0}&\mathbf{0}&\mathbf{0}\\
\mathbf{0}&\mathbf{h}^{[1]}(2)&\mathbf{0}&\mathbf{0}&\mathbf{0}&\mathbf{0}&\mathbf{0}&\mathbf{0}\\
\mathbf{0}&\mathbf{0}&\mathbf{h}^{[1]}(3)&\mathbf{0}&\mathbf{0}&\mathbf{0}&\mathbf{0}&\mathbf{0}\\
\mathbf{0}&\mathbf{0}&\mathbf{0}&\mathbf{h}^{[1]}(1)&\mathbf{0}&\mathbf{0}&\mathbf{0}&\mathbf{0}\\
\mathbf{0}&\mathbf{0}&\mathbf{0}&\mathbf{0}&\mathbf{h}^{[1]}(2)&\mathbf{0}&\mathbf{0}&\mathbf{0}\\
\mathbf{0}&\mathbf{0}&\mathbf{0}&\mathbf{0}&\mathbf{0}&\mathbf{h}^{[1]}(3)&\mathbf{0}&\mathbf{0}\\
\mathbf{0}&\mathbf{0}&\mathbf{0}&\mathbf{0}&\mathbf{0}&\mathbf{0}&\mathbf{h}^{[1]}(1)&\mathbf{0}\\
\mathbf{0}&\mathbf{0}&\mathbf{0}&\mathbf{0}&\mathbf{0}&\mathbf{0}&\mathbf{0}&\mathbf{h}^{[1]}(2)
\end{array}
\right]_{8 \times 24}
\end{eqnarray*}
\end{small}
Then, the received signal at user 1 is
\begin{small}
\begin{eqnarray}\label{eqn:rxsignal3by2}
\mathbf{Y}^{[1]}=
\underbrace{\left[\begin{array}{cc}
\mathbf{h}^{[1]}(1)&\mathbf{0}\\
\mathbf{h}^{[1]}(2)&\mathbf{0}\\
\mathbf{h}^{[1]}(3)&\mathbf{0}\\
\mathbf{0}&\mathbf{h}^{[1]}(1)\\
\mathbf{0}&\mathbf{h}^{[1]}(2)\\
\mathbf{0}&\mathbf{h}^{[1]}(3)\\
\mathbf{0}&\mathbf{0}\\
\mathbf{0}&\mathbf{0}
\end{array}\right]}_{\textrm{rank}=6}
\left[\begin{array}{c}
u^{[1]}_1\\
u^{[1]}_2\\
\vdots\\
u^{[1]}_6
\end{array}\right]
+
\underbrace{\left[
\begin{array}{cc}
\mathbf{h}^{[1]}(1)&\mathbf{0}\\
\mathbf{0}&\mathbf{h}^{[1]}(2)\\
\mathbf{0}&\mathbf{0}\\
\mathbf{h}^{[1]}(1)& \mathbf{0}\\
\mathbf{0}&\mathbf{h}^{[1]}(2)\\
\mathbf{0}&\mathbf{0}\\
\mathbf{h}^{[1]}(1)&\mathbf{0}\\
\mathbf{0}&\mathbf{h}^{[1]}(2)
\end{array}
\right]}_{\textrm{rank}=2}
\left[\begin{array}{c}
u^{[2]}_1\\
u^{[2]}_2\\
\vdots\\
u^{[2]}_6
\end{array}\right]
+
\left[
\begin{array}{c}
z^{[1]}(1)\\
z^{[1]}(2)\\
\vdots\\
z^{[1]}(8)
\end{array}
\right]
\end{eqnarray}
\end{small}
It can be easily seen that the interference is aligned into two
dimensions, spanned by two columns $[1~0~0~1~0~0~1~0]^T$ and
$[0~1~0~0~1~0~0~1]^T$. For the desired signals, since each block
column has rank 3 and two block columns are orthogonal to each
other, the total rank is equal to 6.  It remains to check that the 6
dimensions occupied by the desired signal are linearly independent
of the two interference dimensions. This is true due to the
orthogonality among signals in the seventh and eighth dimensions.
Thus, user 1 is able to achieve 6 DoF. Similar arguments can be used
for user 2, so that 6 DoF are achieved for him as well.

So far, we have seen two problems that we need to consider  and the
ideas used to solve them.  One is the alignment problem and the
other is the linear independence issues. The alignment problem can
be solved by constructing non-overlapping alignment blocks for each
user. Linear independence issues include linear independence of the
desired signals at the desired receiver, and their  independence
with the interference. For the desired signals, we already see that
the signals transmitted over one alignment block are  linearly
independent. The linear independence of data streams transmitted
across different alignment blocks is guaranteed by orthogonality
among alignment blocks in time. For the linear independence between
desired signals and the interference, it is ensured by orthogonality
of signals in the last symbol of each alignment block.  Notice that
in the last symbol of each alignment block, only the signal sent
over this alignment block is active. Thus, after being aligned into
one dimension, the interference vector is linearly independent with
all other signals.

\begin{figure}[!t]
\centering
\includegraphics[width=5.2in, trim=0 210 0 100]{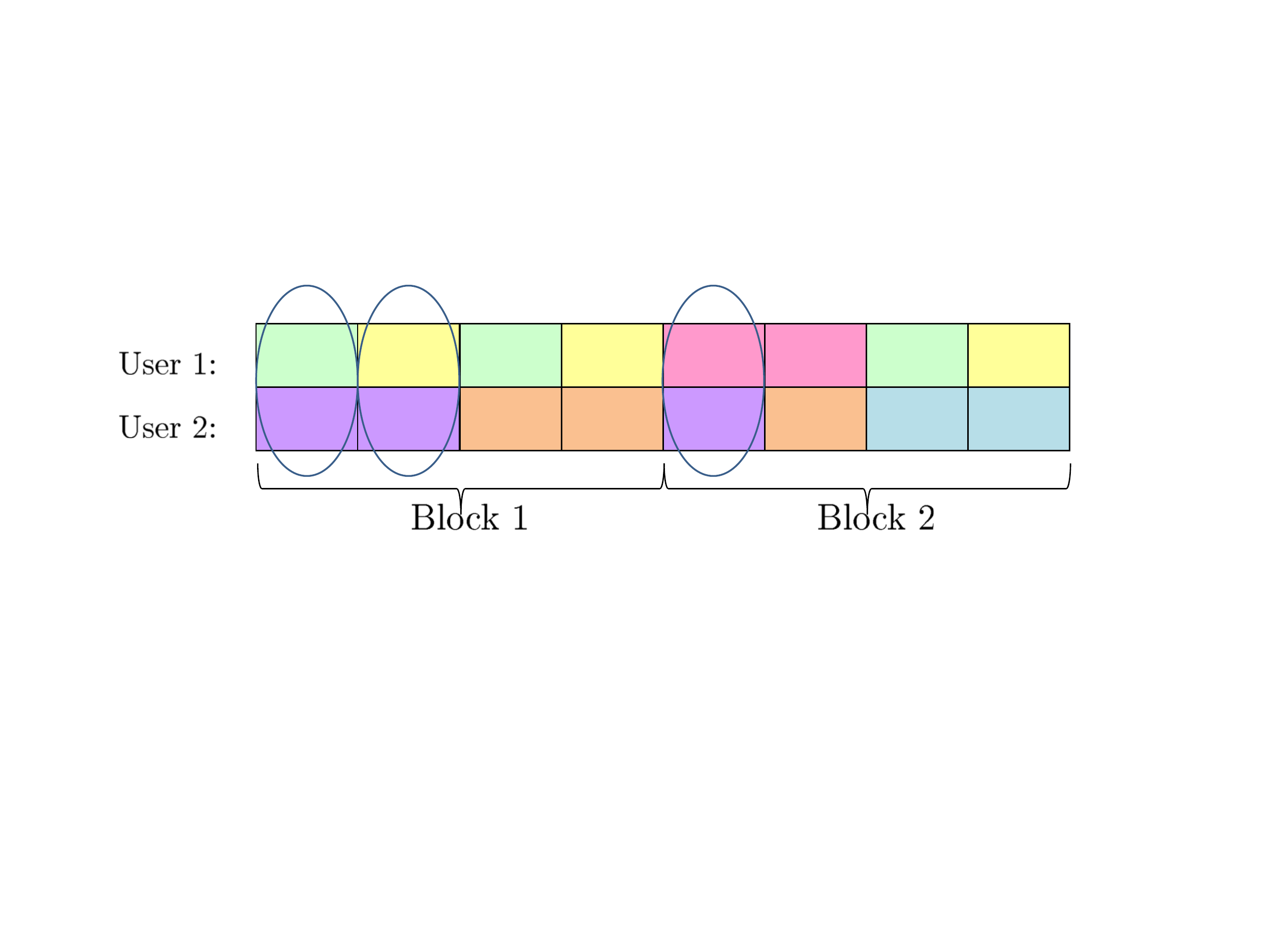}
\caption{The supersymbol structure after reordering for 2 user $3\times 1$
MISO BC} \label{fig:3by2supersymbolreorder}
\end{figure}

To make things systematic, we can change the order of symbols in the
supersymbol and their corresponding rows in the beamforming matrix.
The goal is to {\em separate} the alignment problem and linear
independence issues. In particular, we group last symbols of all
alignment blocks into a block which we refer to as Block 2 as
illustrated in Figure \ref{fig:3by2supersymbolreorder}. The
remaining part is called Block 1. Basically, Block 1 ensures
alignment, while Block 2 guarantees  desired signals do not overlap
with interference. After reordering, the corresponding beamforming
matrix for 2 user $3\times 1$ case is as follows:
\begin{eqnarray}
\left[\begin{array}{cccc}
\mathbf{I}&\mathbf{0}&\mathbf{I}&\mathbf{0}\\
\mathbf{I}&\mathbf{0}&\mathbf{0}&\mathbf{I}\\
\mathbf{0}&\mathbf{I}&\mathbf{I}&\mathbf{0}\\
\mathbf{0}&\mathbf{I}&\mathbf{0}&\mathbf{I} \\
\mathbf{I}&\mathbf{0}&\mathbf{0}&\mathbf{0}\\
\mathbf{0}&\mathbf{I} &\mathbf{0}&\mathbf{0}\\
\mathbf{0}&\mathbf{0}&\mathbf{I}&\mathbf{0}\\
\mathbf{0}&\mathbf{0}&\mathbf{0}&\mathbf{I}
\end{array}\right]
\end{eqnarray}

We can argue that once Block 1 is designed, Block 2 can be
determined automatically. Note every symbol in Block 2 can be
grouped with two symbols in Block 1 as an alignment block. In
addition, over an alignment block, the channel state of the desired
user changes while that of all undesired users remains fixed. Thus,
for the desired user, the last symbol of the alignment block is set
to be the third channel value. For all other undesired users'
channels, they are set to be the same as the channel in previous
symbols of that alignment block. This is illustrated in Figure
\ref{fig:3by2supersymbolreorder} where three symbols constituting
one alignment block for user 1 are circled. Notice that once the two
symbols are determined in Block 1, the third one in Block 2 is
determined uniquely. As a result, {\em we now only need to design
the channel structure in Block 1. Once designed, Block 2 can be
determined automatically.}

\subsubsection{Structure of Block 1 and Design of Beamforming Vectors }

\begin{figure}[!t]
\centering
\includegraphics[width=5.3in, trim=0 200 0 90]{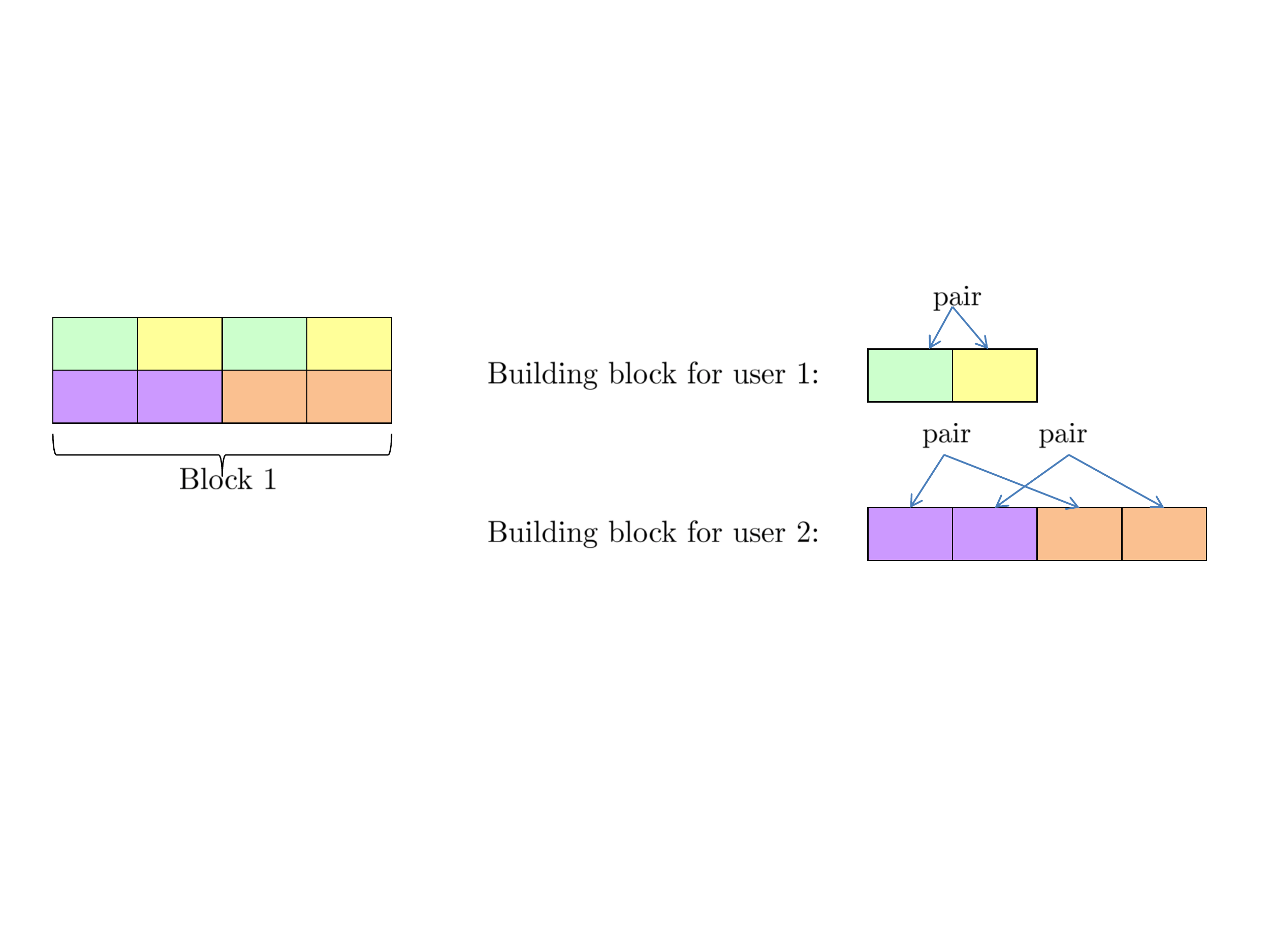}
\caption{Block 1 for 2 user $3\times 1$ MISO BC}
\label{fig:block1for2user}
\end{figure}

In this section, we will show how to design Block 1 of the
supersymbol for $K$ user $3 \times 1$ MISO BC and how to design the
beamforming vectors at the transmitter based on Block 1. To
understand the structure of Block 1, we first consider the 2 user $3
\times 1$ MISO BC. Block 1 is shown in Figure
\ref{fig:block1for2user}. First note that in Block 1, since there is
no third symbol of the alignment block, only two different channel
values for each user are needed. In addition, symbols in Block 1 are
{\em periodic} with the building blocks shown in Figure
\ref{fig:block1for2user}. As we can see, the building block of user
1 consists of $2^1$ symbols while that of user 2  consists of $2^2$
symbols. Since Block 1 consists of 4 symbols, user 1 has two
building blocks while user 2 has only one building block. To design
the beamforming vectors, we can pair the time slots with two
different channel values {\em within every building block} of each
user as shown in Figure \ref{fig:block1for2user} to constitute the
first two symbols of the alignment block .

\begin{figure}[!t]
\centering
\includegraphics[width=5.3in, trim=0 100 0 50]{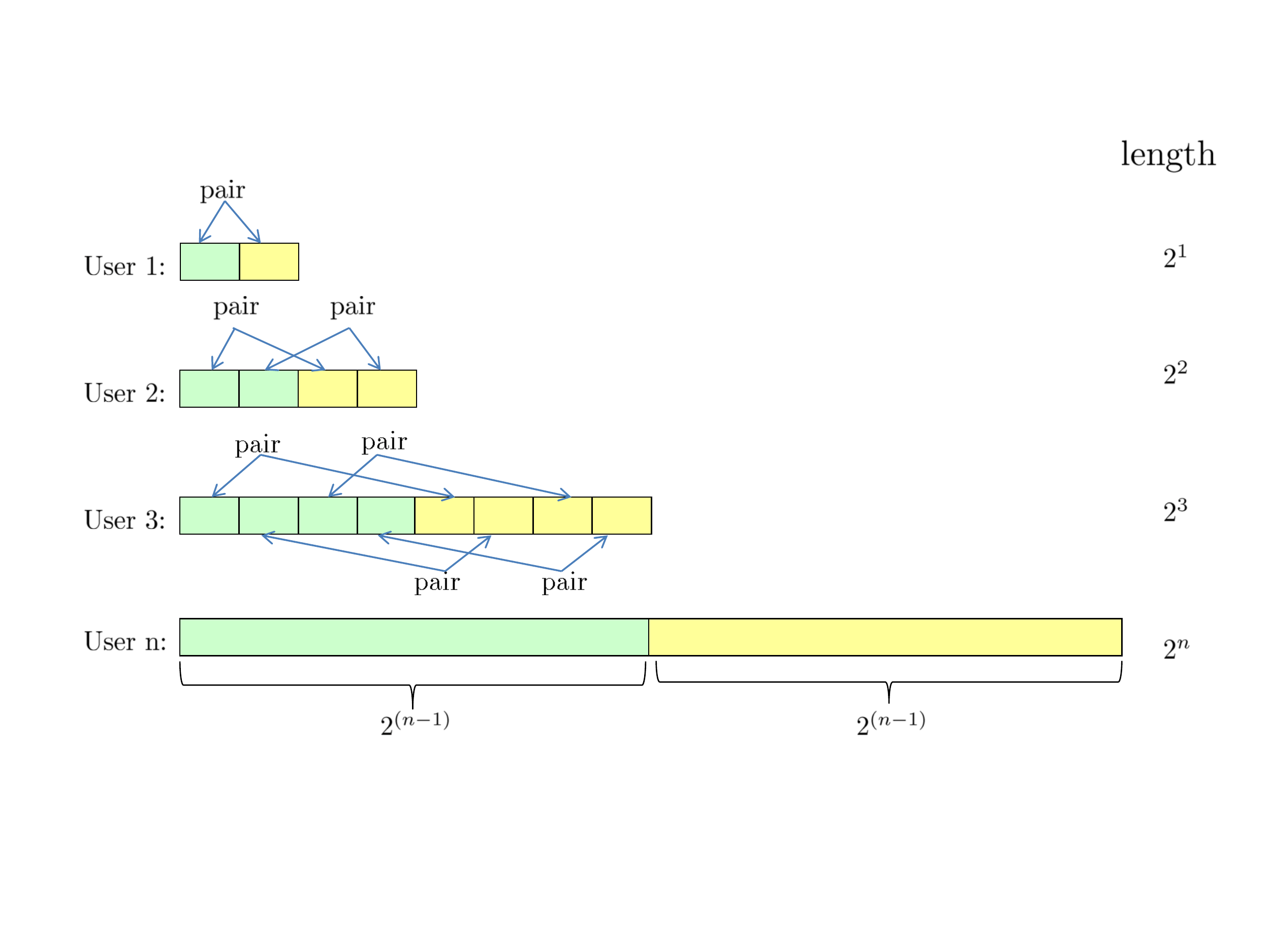}
\caption{Building Blocks for $K$ User $3\times 1$ MISO BC}
\label{fig:buildingblockfor3antenna}
\end{figure}

In general, if there are $K$ users, then Block 1 consists of $2^K$
symbols. Each user's channel states are periodic in Block 1 with the
building block shown in Figure \ref{fig:buildingblockfor3antenna}.
Note that for simplicity, we use the same color to denote the
channel for different users. However, this does not mean that
channel values at different users are the same. In fact, they are
different with probability one.  As we can see, the building block
of user $n$ is comprised of two sub-blocks, each with length
$2^{n-1}$, for a total of $2^n$ symbols. The channel state remains
fixed within each sub-block while it changes across different
sub-blocks. The channel state in the $i^{th}$, $i=1,2$, sub-block
corresponds to the channel vector associated with the $i^{th}$
preset antenna mode at the user. To obtain the temporal correlation
signature for user $n$ in Block 1, we can simply repeat its building
block $2^{K-n}$ times. In other words, there are $2^{K-n}$ building
blocks for user $n$ in Block 1.

To design the beamforming vectors for user $n$, we can pair the
$i^{th}$, $i=1,\cdots,2^{n-1} $ symbol in one sub-block with the one
in the other sub-block {\em within} one building block as shown in
Figure \ref{fig:buildingblockfor3antenna}. Since there are $2^{K-n}$
building blocks, a total of $2^{K-1}$ pairs can be created. It can
be verified these pairs satisfy the requirement of the alignment
block. Detailed explanations are deferred to next section where the
general $K$ user $M \times 1$ MISO BC is considered.

With this construction, we can calculate the achievable DoF. First
let us consider the number of symbols in Block 2. In Block 1, each
user has $2^{K-1}$ pairs. Since each pair needs one symbol in Block
2 to constitute one alignment block, a total of $K\times 2^{K-1}$
symbols are needed in Block 2. Therefore, the total number of
symbols in the supersymbol is $2^K+K\times 2^{K-1}$. On the other
hand, 3 beams can be transmitted over one alignment block for the
desired user while aligned into one dimension at all other users.
Since each user has $2^{K-1}$ alignment block, at the $2^K+K\times
2^{K-1}$ dimensional receiver's signal space,  $3\times 2^{K-1}$
dimensions are occupied by the desired signals while the remaining
$(K-1)\times 2^{K-1}$ dimensions are occupied by the interference.
Since signals are orthogonal in Block 2, desired signals do not
overlap with the interference. Therefore, the normalized DoF is
$\frac{K\times 3\times 2^{K-1}}{2^K+K\times
2^{K-1}}=\frac{3K}{K+2}$.

Next, we take the 3 user $3\times 1$ MISO BC as an example to
illustrate the points mentioned in this section.

\subsubsection{Example: 3 User $3 \times 1$ MISO BC}
\begin{figure}[!h]
\centering
\includegraphics[width=5.3in, clip=true, trim=0 100 0 100]{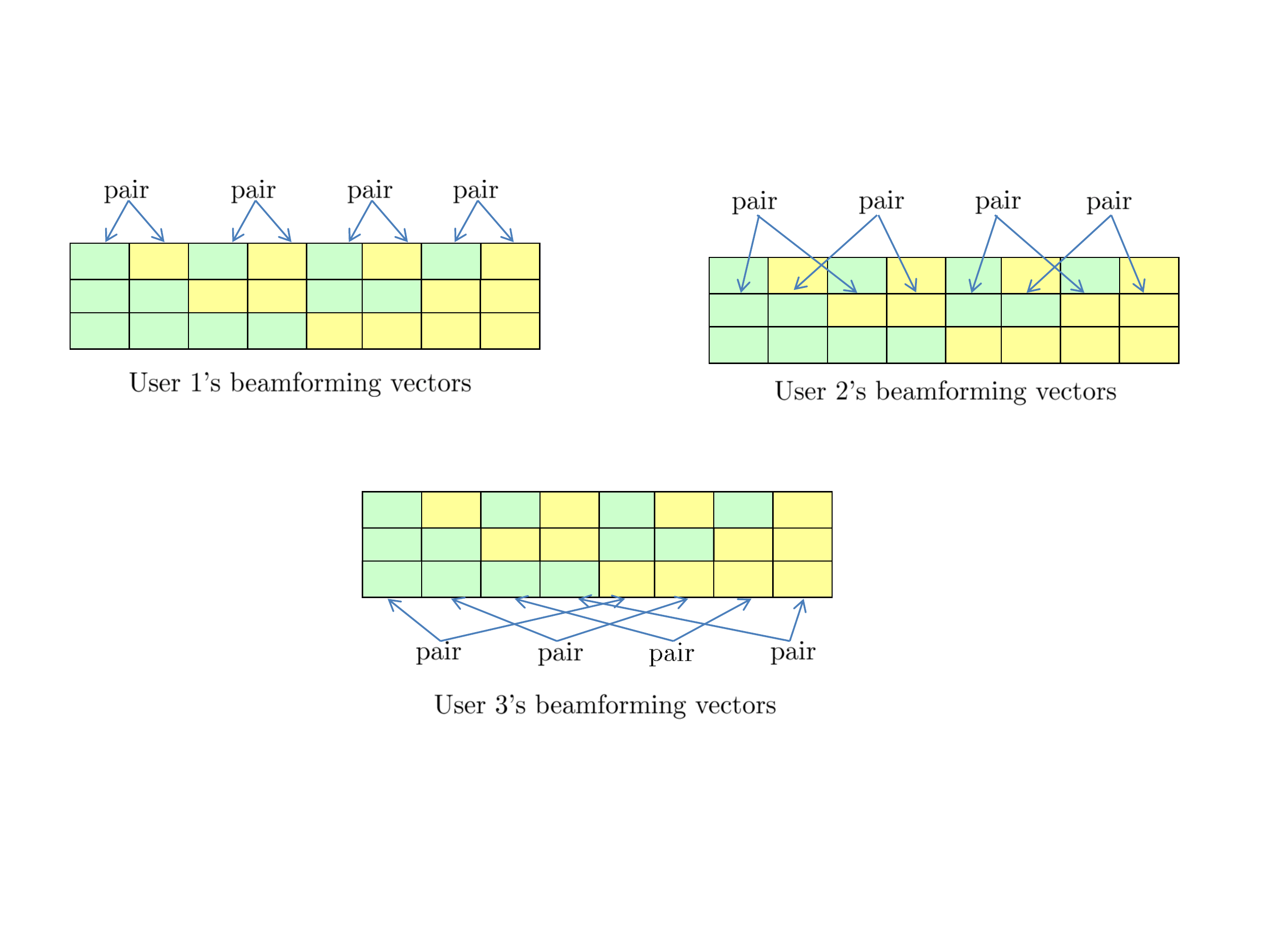}
\caption{Block 1 for 3 User $3\times 1$ MISO BC}
\label{fig:block1for3by3}
\end{figure}
\begin{figure}[!h]
\centering
\includegraphics[width=5.3in, clip=true, trim=0 220 0 200]{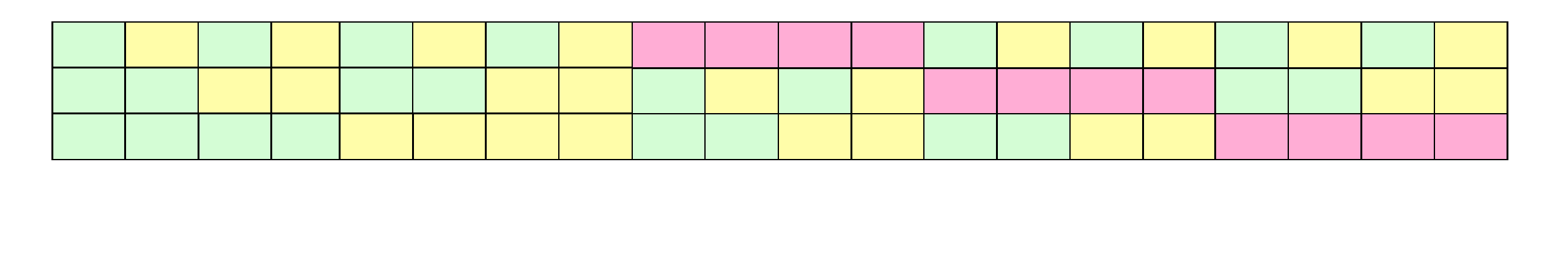}
\caption{The supersymbol structure for 3 user $3\times 1$ MISO BC}
\label{fig:supersymbolfor3by3}
\end{figure}

Let us first consider Block 1. For three users, Block 1 consists of
$2^3$ symbols. The building blocks for user 1, 2 and 3 are shown in
Figure \ref{fig:buildingblockfor3antenna}. Therefore, there are 4, 2
and 1 building blocks for user 1, 2 and 3, respectively. Block 1 is
illustrated in Figure \ref{fig:block1for3by3}, in which the pairing
for each user is also shown.  It can be seen that for each user, the
channel in two time slots of each pair changes at the desired user
while it remains the same at the undesired users, as required by the
alignment block. After adding Block 2, the complete supersymbol
structure for 3 user $3 \times 1$ case is shown in Figure
\ref{fig:supersymbolfor3by3}. The beamforming matrices for user 1, 2
and 3 can be easily obtained through the supersymbol structure as
\begin{small}
\begin{eqnarray*}
\begin{allowdisplaybreaks}
\underbrace{\left[\begin{array}{cccc}
\mathbf{I}&\mathbf{0}&\mathbf{0}&\mathbf{0}\\
\mathbf{I}&\mathbf{0}&\mathbf{0}&\mathbf{0}\\
\mathbf{0}&\mathbf{I}&\mathbf{0}&\mathbf{0}\\
\mathbf{0}&\mathbf{I}&\mathbf{0}&\mathbf{0} \\
\mathbf{0}&\mathbf{0}&\mathbf{I}&\mathbf{0}\\
\mathbf{0}&\mathbf{0} &\mathbf{I}&\mathbf{0}\\
\mathbf{0}&\mathbf{0}&\mathbf{0}&\mathbf{I}\\
\mathbf{0}&\mathbf{0}&\mathbf{0}&\mathbf{I}\\
\mathbf{I}&\mathbf{0}&\mathbf{0}&\mathbf{0}\\
\mathbf{0}&\mathbf{I}&\mathbf{0}&\mathbf{0}\\
\mathbf{0}&\mathbf{0}&\mathbf{I}&\mathbf{0}\\
\mathbf{0}&\mathbf{0}&\mathbf{0}&\mathbf{I}\\
\mathbf{0}&\mathbf{0}&\mathbf{0}&\mathbf{0}\\
\mathbf{0}&\mathbf{0}&\mathbf{0}&\mathbf{0}\\
\mathbf{0}&\mathbf{0}&\mathbf{0}&\mathbf{0}\\
\mathbf{0}&\mathbf{0}&\mathbf{0}&\mathbf{0}\\
\mathbf{0}&\mathbf{0}&\mathbf{0}&\mathbf{0}\\
\mathbf{0}&\mathbf{0}&\mathbf{0}&\mathbf{0}\\
\mathbf{0}&\mathbf{0}&\mathbf{0}&\mathbf{0}\\
\mathbf{0}&\mathbf{0}&\mathbf{0}&\mathbf{0}\\
\end{array}\right]}_{\textrm{user 1}}~~~~~~
\underbrace{
\left[\begin{array}{cccc}
\mathbf{I}&\mathbf{0}&\mathbf{0}&\mathbf{0}\\
\mathbf{0}&\mathbf{I}&\mathbf{0}&\mathbf{0}\\
\mathbf{I}&\mathbf{0}&\mathbf{0}&\mathbf{0}\\
\mathbf{0}&\mathbf{I}&\mathbf{0}&\mathbf{0} \\
\mathbf{0}&\mathbf{0}&\mathbf{I}&\mathbf{0}\\
\mathbf{0}&\mathbf{0} &\mathbf{0}&\mathbf{I}\\
\mathbf{0}&\mathbf{0}&\mathbf{I}&\mathbf{0}\\
\mathbf{0}&\mathbf{0}&\mathbf{0}&\mathbf{I}\\
\mathbf{0}&\mathbf{0}&\mathbf{0}&\mathbf{0}\\
\mathbf{0}&\mathbf{0}&\mathbf{0}&\mathbf{0}\\
\mathbf{0}&\mathbf{0}&\mathbf{0}&\mathbf{0}\\
\mathbf{0}&\mathbf{0}&\mathbf{0}&\mathbf{0}\\
\mathbf{I}&\mathbf{0}&\mathbf{0}&\mathbf{0}\\
\mathbf{0}&\mathbf{I}&\mathbf{0}&\mathbf{0}\\
\mathbf{0}&\mathbf{0}&\mathbf{I}&\mathbf{0}\\
\mathbf{0}&\mathbf{0}&\mathbf{0}&\mathbf{I}\\
\mathbf{0}&\mathbf{0}&\mathbf{0}&\mathbf{0}\\
\mathbf{0}&\mathbf{0}&\mathbf{0}&\mathbf{0}\\
\mathbf{0}&\mathbf{0}&\mathbf{0}&\mathbf{0}\\
\mathbf{0}&\mathbf{0}&\mathbf{0}&\mathbf{0}
\end{array}\right]
}_{\textrm{user 2}}~~~~~~
\underbrace{\left[\begin{array}{cccc}
\mathbf{I}&\mathbf{0}&\mathbf{0}&\mathbf{0}\\
\mathbf{0}&\mathbf{I}&\mathbf{0}&\mathbf{0}\\
\mathbf{0}&\mathbf{0}&\mathbf{I}&\mathbf{0}\\
\mathbf{0}&\mathbf{0}&\mathbf{0}&\mathbf{I} \\
\mathbf{I}&\mathbf{0}&\mathbf{0}&\mathbf{0}\\
\mathbf{0}&\mathbf{I} &\mathbf{0}&\mathbf{0}\\
\mathbf{0}&\mathbf{0}&\mathbf{I}&\mathbf{0}\\
\mathbf{0}&\mathbf{0}&\mathbf{0}&\mathbf{I}\\
\mathbf{0}&\mathbf{0}&\mathbf{0}&\mathbf{0}\\
\mathbf{0}&\mathbf{0}&\mathbf{0}&\mathbf{0}\\
\mathbf{0}&\mathbf{0}&\mathbf{0}&\mathbf{0}\\
\mathbf{0}&\mathbf{0}&\mathbf{0}&\mathbf{0}\\
\mathbf{0}&\mathbf{0}&\mathbf{0}&\mathbf{0}\\
\mathbf{0}&\mathbf{0}&\mathbf{0}&\mathbf{0}\\
\mathbf{0}&\mathbf{0}&\mathbf{0}&\mathbf{0}\\
\mathbf{0}&\mathbf{0}&\mathbf{0}&\mathbf{0}\\
\mathbf{I}&\mathbf{0}&\mathbf{0}&\mathbf{0}\\
\mathbf{0}&\mathbf{I}&\mathbf{0}&\mathbf{0}\\
\mathbf{0}&\mathbf{0}&\mathbf{I}&\mathbf{0}\\
\mathbf{0}&\mathbf{0}&\mathbf{0}&\mathbf{I}
\end{array}\right]}_{\textrm{user 3}}
\end{allowdisplaybreaks}
\end{eqnarray*}
\end{small}

\subsection{$K$ User $M \times 1$ MISO Broadcast Channel}
In this section, we consider the general $K$ user $M\times 1$ case
and show how to systematically design the supersymbol structure and
the beamforming matrix. This consists of three steps.

{\bf \noindent Step 1: Design Block 1}

\begin{figure}[!t]
\centering
\includegraphics[width=5.3in, trim=0 240 0 110]{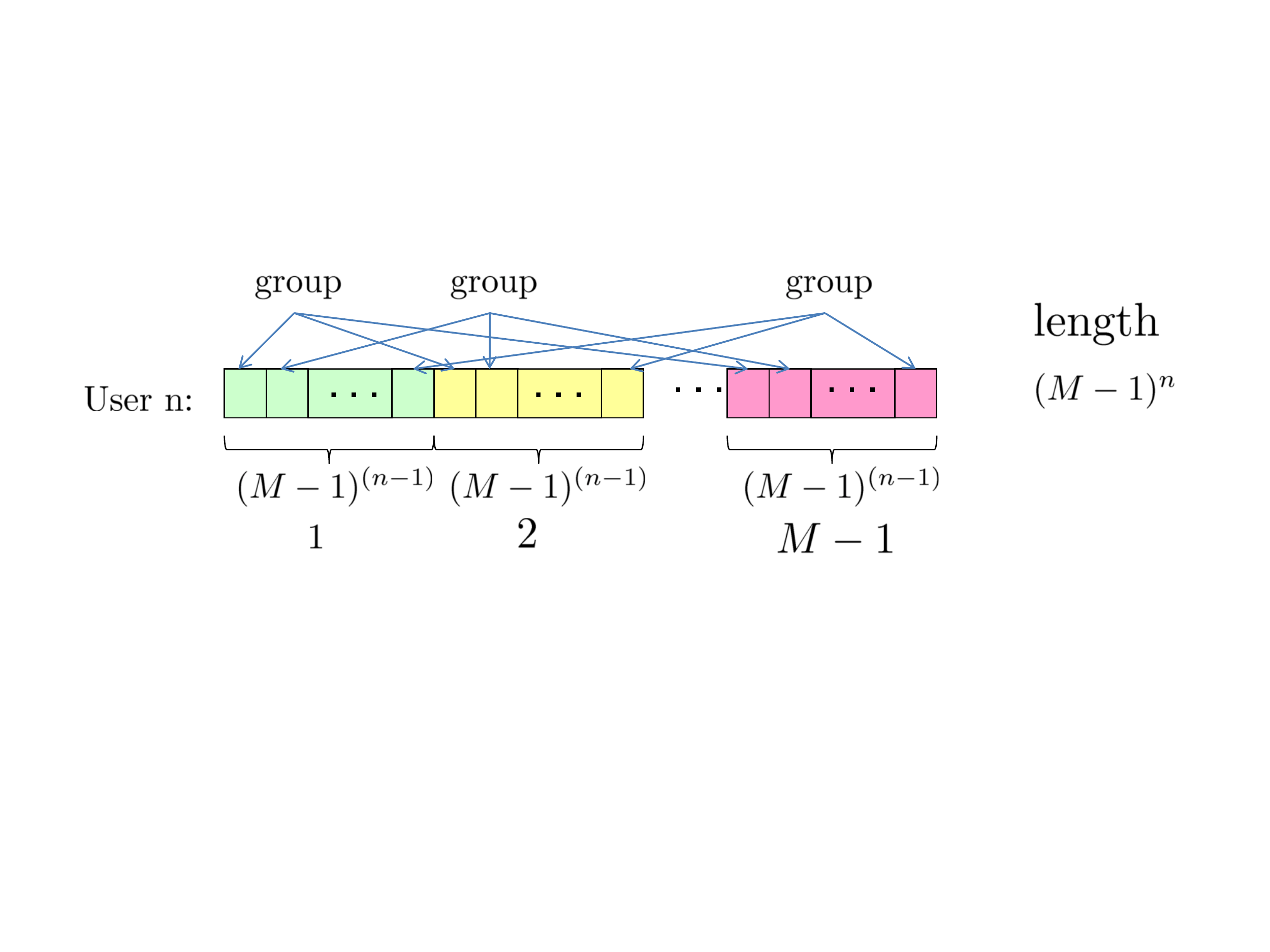}
\caption{The building block of user $n$ for $K$ user $M \times 1$
MISO BC} \label{fig:buildingblockformbyk}
\end{figure}

Block 1 consists of a total of $(M-1)^K$ symbols. Each user's
channel state switching pattern is periodic in Block 1 with the
building block shown in Figure \ref{fig:buildingblockformbyk}. As we
can see, the building block of user $n$ is comprised of $M-1$
sub-blocks, each with length $(M-1)^{n-1}$, for a total of $(M-1)^n$
symbols. The channel state remains fixed within each sub-block while
it changes across different sub-blocks. The channel state in the
$i^{th}$, $i=1,\cdots, M-1$, sub-block corresponds to the channel
vector associated with the $i^{th}$ preset antenna mode at the user. To obtain
the temporal correlation signature for user $n$ in Block 1, we can
simply repeat its building block $(M-1)^{K-n}$ times. In other
words, there are $(M-1)^{K-n}$ building blocks for user $n$ in Block
1.

In general, each user's  temporal correlation can be described by a
function of time whose values come from all the possible channel
values he can take. Let us define the function of time $t\in
\mathbb{N}$ for user $n$ as $f_n(t)$ whose value is drawn from the
set $\{\mathbf{h}^{[n]}(1),\mathbf{h}^{[n]}(2), \cdots,
\mathbf{h}^{[n]}(M)\}$ where $\mathbf{h}^{[n]}(i)$, $i=1,\cdots, M$,
denotes the channel vector associated with the $i^{th}$ mode of user
$n$'s receive antenna.  Then we can denote the channel of Block 1
for user $n$ as
\begin{eqnarray}\label{eqn:Block1}
f_n(t)=\left\{ \begin{array}{cc}
\mathbf{h}^{[n]}(1) & t \equiv 1, 2, \cdots, (M-1)^{n-1} ~~(\!\!\!\!\mod (M-1)^n)\\
\mathbf{h}^{[n]}(2)  & t \equiv (M-1)^{n-1}+1, \cdots, 2(M-1)^{n-1}~ ~(\!\!\!\!\mod (M-1)^n)\\
\vdots & ~  \\
\mathbf{h}^{[n]}(j)  & t \equiv (j-1)(M-1)^{n-1}+1, \cdots, j(M-1)^{n-1} ~~(\!\!\!\!\mod (M-1)^n)\\
\vdots & ~  \\
\mathbf{h}^{[n]}(M-1)  &  t \equiv(M-2)(M-1)^{n-1}+1, \cdots, (M-1)^{n}~~ (\!\!\!\!\mod (M-1)^n)
\end{array}
\right.
\end{eqnarray}
Note that for Block 1, $t=1,2,\cdots, (M-1)^K$, since the length of Block 1 is $(M-1)^K$.

{\bf \noindent Step 2: Design Beamforming Matrices}

With Block 1 designed, we can now design the beamforming vectors. As
mentioned before, the key is to create non-overlapping alignment
blocks for each user. Each alignment block corresponds to one block
column in the beamforming matrix, which is obtained by stacking the
$M \times M$ identity matrix at the rows corresponding to the symbol
instants of the alignment block.

Note that in Block 1, there are $M-1$ distinct symbols for each
user. Therefore, we will construct the first $M-1$ symbols, which
are referred to as a group, of the alignment block. The last symbol
for each alignment block is provided in Block 2. Now consider user
$n$. Recall that the channel state of the desired user changes over
an alignment block. Therefore, we can group the $i^{th}$ symbol in
each of $M-1$ sub-blocks {\em within} one building block as shown in
Figure \ref{fig:buildingblockformbyk}. Since there are $(M-1)^{n-1}$
symbols in one sub-block, a total of $(M-1)^{n-1}$ groups
can be created from one building block. Mathematically, within each
building block, the $i^{th}$ group consists of following symbols:
\begin{eqnarray}
i, i+(M-1)^{n-1},\cdots, i+(M-2)(M-1)^{n-1}~~~i=1,2,\cdots, (M-1)^{n-1}
\end{eqnarray}
Such grouping is repeated within each building block for a total
of $(M-1)^{K-n}$ building blocks for user $n$.  Mathematically, we
can represent the $i^{th}$ group in the $m^{th}$ building block as
\begin{small}
\begin{eqnarray}\label{eqn:group}
(m-1)(M-1)^n+i, (m-1)(M-1)^n+i+(M-1)^{n-1},\cdots, (m-1)(M-1)^n+i+(M-2)(M-1)^{n-1}\notag
\\i=1,2,\cdots, (M-1)^{n-1} ~~ m=1,\ldots, (M-1)^{K-n}
\end{eqnarray}
\end{small}
Note that $(M-1)^n$ is the length of one building block.
\begin{figure}[!t]
\centering
\includegraphics[width=5.3in, clip=true, trim=0 280 0 100]{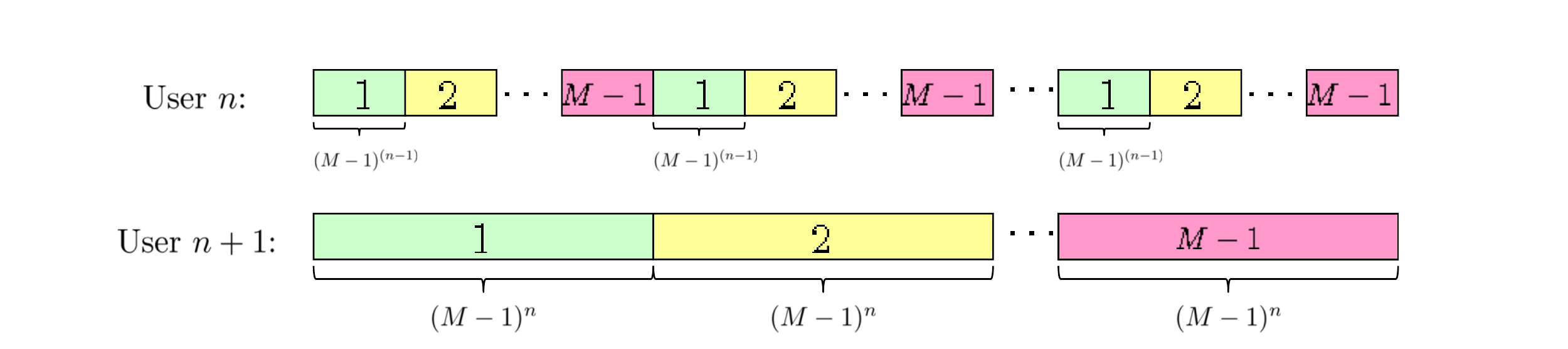}
\caption{The building blocks of user $n$ and $n+1$ for $K$ user $M
\times 1$ MISO BC} \label{fig:buildingblockcompare}
\end{figure}

We need to check that each group satisfies the requirement of the
alignment block, i.e., the channel state of the desired user changes
while that of all other users remains constant in each group.  Note
that \eqref{eqn:group} specifies the time instants of symbols for
each group, and \eqref{eqn:Block1} describes how channel changes
with time for every user. Therefore, we can use \eqref{eqn:Block1}
to verify how channel state changes at each user during the time
instants of each group given in \eqref{eqn:group}. Consider user
$n$. Let us first verify that the channel state changes at the
desired user, i.e., user $n$.  Note that the $i^{th}$ group in the
$m^{th}$ building block is given in \eqref{eqn:group}. Now we
calculate the channel values in these time slots by
\eqref{eqn:Block1}. It can be easily seen that the remainders of
these time slots divided by $(M-1)^n$ are as follows
\begin{eqnarray}\label{eqn:desiredremainder}
i, i+(M-1)^{n-1},\cdots, i+(M-2)(M-1)^{n-1}~~~i=1,2,\cdots, (M-1)^{n-1}.
\end{eqnarray}
Therefore, according to \eqref{eqn:Block1}, the channel values at
these time slots are $\mathbf{h}^{[n]}(1),
\mathbf{h}^{[n]}(2),\cdots, \mathbf{h}^{[n]}(M-1)$.  In other words,
the channel state changes as required. Now let us consider user
$j=1,2,\ldots, n-1$. The channel state should remain fixed for each
group. Notice that the remainders of each group's time slots divided
by $(M-1)^j, j=1,\cdots, n-1$ are the same, i.e., $i \mod (M-1)^j$.
From \eqref{eqn:Block1}, if the remainders are the same, the channel
state is fixed. Now consider user $j=n+1,\cdots, K$. Notice that the
length of sub-blocks of user $j$ is $(M-1)^{j-1}$, which is no
smaller than $(M-1)^{n}$, the length of the building block of user
$n$. In addition, channel of user $j$ remains fixed within each
sub-block. Thus, as long as symbols in each group of user $n$ belong
to the same sub-block of user $j$, then the channel state they see
at user $j$ is the same. Recall that each group is created within
every building block of user $n$. Therefore, if the boundary of the
building block of user $n$ is aligned with that of sub-block of user
$j$, then each group is within the same sub-block at user $j$. This
can be easily verified since the length of the sub-block of
user $j$, $(M-1)^{j-1}$, is an integer multiple of that of the
building block of user $n$, $(M-1)^n$. For example, as illustrated
in Figure  \ref{fig:buildingblockcompare}, the building block of
user $n$ is of length $(M-1)^n$ which is equal to the length of one
sub-block of user $n+1$. The channel remains constant within a
sub-block of user $n+1$. Therefore, all groups within a building
block of user $n$ see the same channel value at receiver $n+1$. As a
result, all groups satisfy the requirement of the alignment block
structure.

{\bf \noindent Step 3: Design Block 2}

Once Block 1 is designed, Block 2 can be easily determined.  Recall
that in Block 1, we create the first $M-1$ symbols of the alignment
block. Therefore, in Block 2, each symbol serves as the last symbol
for each group in Block 1 to create one alignment block. Since for
each user, there are $(M-1)^{K-1}$ groups, a total of $K(M-1)^{K-1}$
symbols are needed for $K$ users in Block 2. Now we can determine
the channel values in Block 2. We divide Block 2 into $K$
sub-blocks, one with length $(M-1)^{K-1}$. In sub-block $n$, we
provide the last symbol for user $n$.  Therefore, user $n$'s channels
are equal to $\mathbf{h}^{[n]}(M)$ in sub-block $n$. For all other users, each of $(M-1)^{K-1}$ symbols in sub-block $n$ is set to be equal to the value of its corresponding group. As we have shown
before, each group's channel remains fixed at the undesired user.
Thus, to determine each symbol in sub-block $n$ for user $j=1,\cdots,n-1,n+1,\cdots,K$, we can find the first time instant of its corresponding group in Block 1, then
set it to be equal to the channel value at that time instant. From \eqref{eqn:group},
we can see that the first time instant of the $i^{th}$ group in the
$m^{th}$ building block for user $n$ is
\begin{eqnarray}
t^{[n]}_1=(m-1)(M-1)^n+i~~~~i=1,2,\cdots, (M-1)^{n-1} ~~ m=1,\ldots, (M-1)^{K-n}
\end{eqnarray}
Therefore, user $j=1,\cdots,n-1,n+1,\cdots,K$'s channel in the
$n^{th}$ sub-block in Block 2 is equal to $f_j(t^{[n]}_1)$.

With this scheme, each user achieves $M(M-1)^{K-1}$ DoF over
$(M-1)^{K}+K(M-1)^{K-1}$ symbol extensions. Therefore, the
normalized total DoF is equal to
$\frac{KM(M-1)^{K-1}}{(M-1)^{K}+K(M-1)^{K-1}}=\frac{MK}{M+K-1}$.

{\it Remark:} Note that the beamforming strategy does not require
cooperation among antennas. Thus, the same achievable schemes
derived for $K$ user $M\times 1$ MISO BC can be directly applied to
the $M\times K$ $X$ channel where there are $M$ transmitters and $K$
receivers. Each transmitter has a message for each receiver for a
total of $MK$ messages in the network.

\section{Achievable Rates for the $K$ User $M \times 1$ MISO BC with Zero-Forcing Interference at the Receiver}
In this section, we derive a closed form expression for the rate
achieved using the blind interference alignment scheme  designed in
last section and with zero-forcing interference at the receiver. The
result is presented in the following theorem.
\begin{theorem}
For the $K$ user $M\times 1$ MISO BC defined in Section
\ref{sec:systemmodel}, the achievable sum rate with zero-forcing
interference at each user, is
\begin{eqnarray}
R&=&\sum_{k=1}^{K}\frac{1}{M+K-1}\mathbb{E}\left[\log\det \left(\mathbf{
I}+\frac{(K+M-1)P}{M^2K}\mathbf {H}^{[k]}
\mathbf{H}^{[k]\dag}\right)\right]
\end{eqnarray}
where
\begin{eqnarray}
\mathbf {H}^{[k]}=\left[\begin{array}{c}
\frac{1}{\sqrt{K}} \mathbf{h}^{[k]}(1)\\
\frac{1}{\sqrt{K}} \mathbf{h}^{[k]}(2)\\
\vdots\\
\frac{1}{\sqrt{K}} \mathbf{h}^{[k]}(M-1)\\
\mathbf{h}^{[k]}(M)\\
\end{array}\right]
\end{eqnarray}
\end{theorem}

Since interference alignment is achieved perfectly within a finite
number of symbol extensions, the sum rate provides a capacity
approximation within $\mathcal{O}(1)$. In other words, the
approximation error is bounded by a constant as SNR goes to
infinity. We can approximate the achievable rate at high SNR as
follows.
\begin{eqnarray}
R^{[k]}&\approx&\frac{1}{M+K-1}\mathbb{E}\left[\log\det \left(\frac{(K+M-1)P}{M^2K}\mathbf {H}^{[k]}
\mathbf{H}^{[k]\dag}\right)\right]\notag\\
&=&\frac{1}{M+K-1}\mathbb{E}\left[\log\det \left(\frac{(K+M-1)P}{M^2K}\mathbf {\bar{H}}^{[k]}
\mathbf{\bar{H}}^{[k]\dag}\right)\right]-\frac{(M-1)\log K}{M+K-1}
\end{eqnarray}
where
\begin{eqnarray}
\mathbf {\bar{H}}^{[k]}=\left[\begin{array}{c}
\mathbf{h}^{[k]}(1)\\
\mathbf{h}^{[k]}(2)\\
\vdots\\
\mathbf{h}^{[k]}(M-1)\\
\mathbf{h}^{[k]}(M)\\
\end{array}\right]
\end{eqnarray}

\begin{proof}
We begin with the 2 user $2\times 1$ case. Due to symmetry, let us
consider user 1. The received signal at user 1 is given by
\eqref{eqn:rxsingalatuser1}. The interference is aligned along
$[1~0~1]^T$. To zero-force the interference, we need to project the
received signal onto a 2 dimensional subspace that is orthogonal to
$[1~0~1]^T$. It can be seen that rows of the following matrix form
an orthonormal basis of this subspace.
\begin{eqnarray}
\mathbf{P}=\left[ \begin{array}{ccc}
\frac{1}{\sqrt{2}} & 0 &-\frac{1}{\sqrt{2}}\\
0&1&0
\end{array}\right]
\end{eqnarray}
Thus, after projection, the signal is
\begin{eqnarray}
\mathbf{P}\mathbf{y}^{[1]}=\underbrace{\left[\begin{array}{c}
\frac{1}{\sqrt{2}} \mathbf{h}^{[1]}(1)\\
\mathbf{h}^{[1]}(2)\\
\end{array}\right]}_{\mathbf{H}^{[1]}}\left[\begin{array}{c}u^{[1]}_1\\ u^{[1]}_2\end{array}\right]+\tilde{\mathbf{z}}
\end{eqnarray}
where $\tilde{\mathbf{z}}=\mathbf{P}\mathbf{z}$ is the noise, still
white, after the projection. Thus, receiver 1 accesses a full rank
$2\times 2$ MIMO channel. Essentially,  what the zero forcing
receiver does is using the interference received over the last time
slot of each alignment block to cancel the interference received in
previous time slots in that alignment block.  Since there is no
CSIT, we allocate equal power to each data stream, i.e.,
$\frac{3P}{8}$ per data stream  such that the average power
constraint over three time slots is satisfied. Thus, the achievable
rate for user 1 per time slot is
\begin{eqnarray}
R^{[1]}=\frac{1}{3}\mathbb{E}\left[\log\det \left(\mathbf{
I}+\frac{3P}{8}{\mathbf H}^{[1]}\mathbf{H}^{[1]\dag}\right)\right]
\end{eqnarray}
Due to symmetry, user 2 is able to achieve
\begin{eqnarray}
R^{[2]}=\frac{1}{3}\mathbb{E}\left[\log\det \left(\mathbf{
I}+\frac{3P}{8}{\mathbf H}^{[2]}\mathbf{H}^{[2]\dag}\right)\right]
\end{eqnarray}
where
\begin{eqnarray}
\mathbf{ H}^{[2]}=\left[\begin{array}{c}
\frac{1}{\sqrt{2}} \mathbf{h}^{[2]}(1)\\
\mathbf{h}^{[2]}(2)\\
\end{array}\right]
\end{eqnarray}
For the $K$ user  $2\times 1$ MISO BC, from \eqref{eqn:2byK}, it can
be seen that  at user 1 the subspaces occupied by the desired signal
and interference are spanned by the first two columns and the last
$K-1$ columns of the following matrix, respectively.
\begin{eqnarray}
\left[
\begin{array}{cccc}
\mathbf{h}^{[1]}(1)&1&\cdots&1\\
\mathbf{h}^{[1]}(2)&0&\cdots&0\\
\mathbf{0}&1&\cdots&0\\
\vdots&\vdots&\ddots&\vdots\\
\mathbf{0}&0&\cdots&1
\end{array}
\right]_{(K+1) \times (K+1)}
\end{eqnarray}
It can be seen that the rows of the following matrix form an
orthonormal basis of the subspace orthogonal to the subspace spanned
by interference.
\begin{eqnarray}
\mathbf{P}=\left[ \begin{array}{ccccc}
\frac{1}{\sqrt{K}} & 0 &-\frac{1}{\sqrt{K}}& \cdots &-\frac{1}{\sqrt{K}}\\
0&1&0&\cdots&0
\end{array}\right]_{2\times (K+1)}
\end{eqnarray}
After projection, the signal is
\begin{eqnarray}
\mathbf{y}^{[1]'}=\underbrace{\left[\begin{array}{c}
\frac{1}{\sqrt{K}} \mathbf{h}^{[1]}(1)\\
\mathbf{h}^{[1]}(2)\\
\end{array}\right]}_{\mathbf{H}^{[1]}}\left[\begin{array}{c}u^{[1]}_1\\ u^{[1]}_2\end{array}\right]+\tilde{\mathbf{z}}
\end{eqnarray}
With equal power allocation to each data stream and under the average power constraint, user 1's rate is
\begin{eqnarray}
R^{[1]}=\frac{1}{K+1}\mathbb{E}\left[\log\det \left(\mathbf{
I}+\frac{(K+1)P}{4K}\mathbf {H}^{[1]}
\mathbf{H}^{[1]\dag}\right)\right]
\end{eqnarray}
Due to symmetry, user $k$ can achieve a rate
\begin{eqnarray}
R^{[k]}=\frac{1}{K+1}\mathbb{E}\left[\log\det \left(\mathbf{
I}+\frac{(K+1)P}{4K}\mathbf {H}^{[k]}
\mathbf{H}^{[k]\dag}\right)\right]
\end{eqnarray}
where
\begin{eqnarray}
\mathbf{ H}^{[k]}=\left[\begin{array}{c}
\frac{1}{\sqrt{K}} \mathbf{h}^{[k]}(1)\\
\mathbf{h}^{[k]}(2)\\
\end{array}\right]
\end{eqnarray}

Now let us consider the 2 user $3 \times 1$ case. Different from the
$K$ user $2\times 1$ case, each user has more than one alignment
block. Recall that alignment blocks are orthogonal with each other.
Thus, we can decode signals sent over different alignment blocks
separately. Consider the received signal at user 1 which is given by
\eqref{eqn:rxsignal3by2}. Since two block columns are orthogonal, we
can first decode the signals along the first block column by
discarding three dimensions occupied by the desired signal along the
second block column in the received signal space. Therefore, in the
remaining 5 dimensional subspace of the 8 dimensional signal space,
the desired signals are along the first three columns and the
interference are along the last two columns of the following matrix.
\begin{eqnarray}
\left[
\begin{array}{ccc}
\mathbf{h}^{[1]}(1)&1& 0\\
\mathbf{h}^{[1]}(2)&0& 1 \\
\mathbf{h}^{[1]}(3)&0&0 \\
\mathbf{0}&1&0\\
\mathbf{0}&0&1
\end{array}
\right]_{5 \times 5}
\end{eqnarray}
The projection matrix is
\begin{eqnarray}
\mathbf{P}=\left[ \begin{array}{ccccc}
\frac{1}{\sqrt{2}} & 0 & 0 & -\frac{1}{\sqrt{2}} & 0\\
0 & \frac{1}{\sqrt{2}} & 0 & 0 & -\frac{1}{\sqrt{2}}\\
0 & 0 & 1 & 0 & 0
\end{array}\right]
\end{eqnarray}
After projection, the signal is
\begin{eqnarray}
\mathbf{y}^{[1]'}=\underbrace{\left[\begin{array}{c}
\frac{1}{\sqrt{2}} \mathbf{h}^{[1]}(1)\\
\frac{1}{\sqrt{2}} \mathbf{h}^{[1]}(2)\\
\mathbf{h}^{[1]}(3)\\
\end{array}\right]}_{\mathbf{H}^{[1]}}\left[\begin{array}{c}u^{[1]}_1\\ u^{[1]}_2\\ u^{[1]}_3\end{array}\right]+\tilde{\mathbf{z}}
\end{eqnarray}
Thus, the rate for three data streams over 8 symbol extensions is
\begin{eqnarray}
R^{[1]'}=\mathbb{E}\left[\log\det \left(\mathbf{
I}+\frac{2P}{9}\mathbf {H}^{[1]}
\mathbf{H}^{[1]\dag}\right)\right]
\end{eqnarray}
Due to symmetry, the rate for the other three data streams
transmitted over the other alignment block is the same. Therefore,
2$R^{[1]'}$ can be achieved over 8 symbol extensions. Thus, the
normalized rate is
\begin{eqnarray}
R^{[1]}=\frac{1}{4}\mathbb{E}\left[\log\det \left(\mathbf{
I}+\frac{2P}{9}\mathbf {H}^{[1]}
\mathbf{H}^{[1]\dag}\right)\right]
\end{eqnarray}
By similar argument, we can obtain user 2's rate by changing the user index 1 into 2.

For general $K$ user $M \times 1$ case, to decode $M$ data streams
over one alignment block, we can discard dimensions occupied by all
desired signals transmitted over all other alignment blocks since they are orthogonal in time. For interference, we can use the interference signal received  over the last symbol of each alignment block to cancel interference received over previous symbols in that alignment block. Because the interference is the same over all symbols of each alignment block due to repetition coding and interference is orthogonal over the last symbol of all alignment blocks.  Mathematically, it can be shown that for user $k$, the first block column corresponding to
$M$ columns of the following matrix spans the subspace occupied by the desired signal in one alignment block while
the last $(K-1)(M-1)$ columns span the subspace occupied by the
interference.
\begin{eqnarray}
\left[
\begin{array}{ccccc}
\mathbf{H}^{[k]}_{1:M-1}&\mathbf{I}& \mathbf{I}&\cdots&\mathbf{I}\\
\mathbf{0}_{(M-1)\times M}&\mathbf{I}& \mathbf{0}& \cdots&\mathbf{0} \\
\mathbf{0}_{(M-1)\times M}&\mathbf{0}& \mathbf{I}& \cdots&\mathbf{0} \\
\vdots&\vdots & \vdots & \ddots & \vdots\\
\mathbf{0}_{(M-1)\times M}&\mathbf{0}&\mathbf{0}&\cdots&\mathbf{I}\\
\mathbf{h}^{[k]}(M)&\mathbf{0}_{1\times (M-1)}&\mathbf{0}_{1\times (M-1)}&\mathbf{0}_{1\times (M-1)}&\mathbf{0}_{1\times (M-1)}
\end{array}
\right]_{(KM-K+1)\times \left(M+(K-1)(M-1)\right)}
\end{eqnarray}
where $\mathbf{I}$ is the $(M-1) \times (M-1)$ identity matrix, $\mathbf{0}$ is the $(M-1) \times (M-1)$  zero matrix, and
\begin{eqnarray}
\mathbf{H}^{[k]}_{1:M-1}=\left[\begin{array}{c}
 \mathbf{h}^{[k]}(1)\\
\mathbf{h}^{[k]}(2)\\
\vdots\\
 \mathbf{h}^{[k]}(M-1)\\
\end{array}\right]
\end{eqnarray}
Thus, the projection matrix is
\begin{eqnarray}
\mathbf{P}=\left[ \begin{array}{cccccc}
\frac{1}{\sqrt{K}}\mathbf{I}&-\frac{1}{\sqrt{K}}\mathbf{I}&-\frac{1}{\sqrt{K}}\mathbf{I}&\cdots&-\frac{1}{\sqrt{K}}\mathbf{I}&0\\
\mathbf{0}_{1\times (M-1)}&\mathbf{0}_{1\times (M-1)}&\mathbf{0}_{1\times (M-1)}&\cdots&\mathbf{0}_{1\times (M-1)}&1\\
\end{array}
\right]_{M\times (KM-K+1)}
\end{eqnarray}
Then the received signal after zero forcing is
\begin{eqnarray}
\mathbf{y}^{[k]'}=\underbrace{\left[\begin{array}{c}
\frac{1}{\sqrt{K}} \mathbf{h}^{[k]}(1)\\
\frac{1}{\sqrt{K}} \mathbf{h}^{[k]}(2)\\
\vdots\\
\frac{1}{\sqrt{K}} \mathbf{h}^{[k]}(M-1)\\
\mathbf{h}^{[k]}(M)\\
\end{array}\right]}_{\mathbf{H}^{[k]}}\left[\begin{array}{c}u^{[k]}_1\\ u^{[k]}_2\\ \vdots\\ u^{[k]}_M\end{array}\right]+\tilde{\mathbf{z}}
\end{eqnarray}
With equal power allocation to each data stream, the rate achieved for $M$ data streams in one alignment block is
\begin{eqnarray}
R^{[k]'}=\mathbb{E}\left[\log\det \left(\mathbf{
I}+\frac{(K+M-1)P}{M^2K}\mathbf {H}^{[k]}
\mathbf{H}^{[k]\dag}\right)\right]
\end{eqnarray}
Since there are a total of $(M-1)^{K-1}$ alignment blocks, and a
total of $(M-1)^{K}+K(M-1)^{K-1}$ time slots, the normalized rate
for user $k$ is
\begin{eqnarray}
R^{[k]}&=&\frac{(M-1)^{K-1}}{(M-1)^{K}+K(M-1)^{K-1}}\mathbb{E}\left[\log\det \left(\mathbf{
I}+\frac{(K+M-1)P}{M^2K}\mathbf {H}^{[k]}
\mathbf{H}^{[k]\dag}\right)\right]\\
&=&\frac{1}{M+K-1}\mathbb{E}\left[\log\det \left(\mathbf{
I}+\frac{(K+M-1)P}{M^2K}\mathbf {H}^{[k]}
\mathbf{H}^{[k]\dag}\right)\right]
\end{eqnarray}

\end{proof}

\section{Conclusion}

Recent work has shown that channel correlations can be
exploited to achieve interference alignment even when the
transmitter has no information about the exact channel values
\cite{Jafar_SCBC}. This work shows that channel temporal
correlations required in \cite{Jafar_SCBC} can be created, and thus
interference alignment can be achieved in practice. The idea is to
manipulate channels through antenna selection. By switching antennas
during transmission, different temporal correlations can be created
at different users. With these new insights, we provide a systematic
way to achieve blind interference alignment for the $K$ user $M \times 1$ MISO BC, possibly with
multicast traffic so that the outer bound of $\frac{MK}{M+K-1}$ DoF is achieved. The coding scheme is essentially a simple repetition code over a finite number of symbols. The key to the blind interference alignment is that multiple symbols that follow the same repetition code are automatically aligned into one dimension and can only be separated through short term channel fluctuations. Thus the key is to introduce these short term channel fluctuations at the correct time instants to un-align the desired symbols without disturbing the alignment of the interfering symbols. The simplicity of the coding scheme allows us to write closed form achievable rate expressions at any SNR.  While the assumption of perfect CSIR allows us to use the DoF metric and write closed form rate expressions, CSIR is not not needed to cancel interference. This feature makes the schemes proposed in this work suitable for non-coherent communications, i.e., with no CSIR,  in combination with differential coding schemes that require coherence over relatively short intervals.

In this paper we focus only on blind interference alignment schemes that exploit
\emph{receive} antenna selection. This does not imply that transmit
antenna selection is never useful. Indeed, the two user MIMO
interference channel blind interference alignment example presented
in \cite{Jafar_SCBC} is easily seen to be possible through  antenna
switching only at one \emph{transmitter}. In addition, we only
consider the broadcast channel and $X$ channel in this paper. In
fact, antenna switching can be useful in other networks as well. For
example, with antenna switching, the $K$ user interference
channel with constant channel coefficients can achieve more than one DoF. For example, the 4 user interference channel has at least $\frac{4}{3}$ DoF using configurable antenna at the receiver. This can be seen by converting the $2\times 2$  $X$ channel into 2 interfering MAC channels. In each MAC channel there are two transmitters each sending one message to the receiver and causing interference to the receiver in the other MAC channel. Now, replicate each receiver twice to create a receiver for each transmitter, these receivers are statistically equivalent, hence each is
equivalent to the MAC receiver. As a result, it is converted into a 4 user interference channel that achieves  $\frac{4}{3}$ DoF. Exploiting SCBC schemes for other wireless network is also an interesting avenue for future work.

\end{document}